\documentclass[a4paper]{article}

\usepackage{amsfonts}
\usepackage{epic}
\usepackage{epsfig}
\usepackage{subfigure}
\usepackage{amsmath}
\usepackage{amssymb}
\usepackage{rotating}
\usepackage{curves}
\begin{document}

\title{\bf Quantum properties of a cyclic structure based on tripolar fields}

\author{V. N. Yershov \\
{\small \it Mullard Space Science Laboratory (University College London),} \\ 
{ \small \it  Holmbury St.Mary, Dorking RH5 6NT, UK} \\
{ \small vny@mssl.ucl.ac.uk}}

\date{}

\maketitle

\begin{abstract}
The properties of cyclic structures (toroidal oscillators) 
based on classical tripolar (colour) fields are discussed, in particular, 
of a cyclic structure formed of three colour-singlets spinning around a 
ring-closed axis. It is shown that the helicity and handedness of 
this structure can be related to the quantum properties of the electron. 
The symmetry of this structure corresponds to the complete cycle 
of $\frac{2}{3}\pi$-rotations of its constituents, which leads to 
the exact overlapping of the paths of its three complementary 
coloured constituents, making the system dynamically colourless. 
The gyromagnetic ratio of this system is estimated to be 
g$\approx 2$, which agrees with the Land\'e g-factor for the electron.
\end{abstract}

\newcommand{\mybox}[1]{{\small \vbox{\hrule 
\hbox{\vrule\,\strut #1\,\vrule}\hrule}}}

\newcommand{\abs}[1]{\lvert#1\rvert}
\newcommand{\DoT}[1]{\begin{turn}{-180}\raisebox{-1.67ex}{#1}\end{turn}}
\newcommand{\up}{\upharpoonright}
\newcommand{\down}{\downharpoonright}
\newcommand{\DN}{\hspace{-0.12em}
{\DoT{\DoT{$\triangle$}}}\hspace{-0.7em}\raisebox{0.3ex}
{\tiny \DoT{\DoT{$-$}}}\hspace{0.15em}}
\newcommand{\DP}{\hspace{-0.12em}
{\DoT{\DoT{$\triangle$}}}\hspace{-0.7em}\raisebox{0.3ex}
{\tiny \DoT{\DoT{$+$}}}\hspace{0.15em}
\hspace{0.2ex}}

\section{Introduction}

It is widely accepted that quantum and classical mechanics, despite
being conceptually different, are intimately related 
\cite{landau81}, which can be seen, for instance, in the necessity to 
describe any quantum object in the context of a classical 
system (measuring device). 
Practical applications of quantum 
mechanics, e.g., in quantum computing, also show that quantum 
information can only be transmitted in conjunction with classical 
signalling \cite{penrose98}.
There exist many similarities 
between classical and quantum phenomena, although they occur 
in completely different contexts \cite{elitzur05}. 
This leads to the possibility of describing deterministic systems 
with the use of quantum-mechanical formalism and vice-versa, thus 
arriving at a deeper understanding of irreversibility, causality and 
unpredictability concepts, as was shown by I.\,Prigogine \cite{prigogine62}
and others \cite{strocchi66}. 
Nowadays, exploring the mechanisms responsible for the appearance 
of a classical world through decoherence of quantum systems 
is regarded as one of the most important tasks of quantum mechanics 
\cite{joos03}. For instance, it is known that a regular 
pattern could emerge as a result of interactions 
between purely chaotic systems \cite{almeida05}. 
A series of models showing the possibility of mapping the quantum 
states of a system onto the states of a completely deterministic 
model were also discussed recently by \,'t~Hooft \cite{hooft01a}, 
Prezhdo \cite{prezhdo02}, and others 
\cite{keshavamurthy97}.

On the other hand, there are models showing that a classical 
system can manifest itself quantum-mechanically  \cite{hooft97},
which means that quantum field theories could be underlied by 
classical mechanics \cite{elze05}, with  quantum uncertainties 
also having a deterministic origin \cite{durr92}.  
Many physicists explore the possibility that quantum phenomena 
could arise as a result of information loss due to non-reversible 
dissipative processes and self-organisation in nonlinear deterministic 
systems \cite{prigogine01}. 

Here we shall follow this lead and examine
two composite classical systems with nonlinear 
pairwised potentials, showing that such systems
could exhibit quantum properties identifiable 
with those of the electron and its neutrino. Today's 
commonly-held view is that these particles are point-like 
quantum objects whose classical description is impossible 
because the microscopic reality is controlled by non-commuting 
operators, and the more so because quantum models are able to
account for most of the experimental data \cite{griffiths87}. 
However, Dirac, the creator of quantum theory of 
the electron, warned that his point-electron model was 
actually a mathematical approximation, not conforming to current 
physical ideas \cite{dirac39}. The point model of a charged 
particle is physically unstable and requires the density of the 
particle's rest mass-energy to be infinite. According to the laws of 
electrodynamics, a point charge would have to have a zero spin 
and zero magnetic moment.  Thus, the laws of magnetism dictate 
that the electron {\it must} have some physical extent to have 
a magnetic moment. That is why modern quantum field theories  
do smear the charge of the physical electron over some 
extended region in order to renormalise the theory and produce 
finite results \cite{bernabeu00}. 

Our model is based on classical tripolar (colour) fields
with the sources of these fields -- tripolar charges -- 
regarded as primitive entities with almost no properties, 
except for those determined by the properties of space. 
Therefore, these entities are likely to be described
as autosolitons -- localised eigenstates of the manifold,
which have the properties of both particles and waves,
propagate and interact with each other and obey the energy
and momentum conservation laws \cite{kerner94}. The autosoliton 
parameters are entirely determined by the parameters of the system 
and do not depend on the excitation causing their formation.   
The matter particles could then be seen as composite systems 
based on stable configuration patterns of a moving manifold 
(space), which is not a novelty: perhaps the first 
who suggested that elementary particles might be organised
patterns of space was Wheeler \cite{wheeler62}. 
There are also many other models of this kind \cite{bohm80}.
An important feature to be taken into account in this framework 
is torsion of the manifold \cite{ritis83}, which would 
lead to the nonlinear Heisenberg equation \cite{dzhunushaliev98} 
and non-Abelian degrees of freedom. The corresponding field 
would have a topological quantum number -- the colour analogy of 
helicity in fluid dynamics \cite{jackiw00} leading to colour
solitons \cite{dai06}. 
By assuming an appropriate interaction potential between colour 
particles in a many-particles system (e.g., of the Lennard-Jones 
type, with the long-range attractive and short-range 
repulsive character) one can reveal a potential surface with 
multiple local minima leading to kinematic constraints
of a topological nature.
This is entirely analogous to the cluster formation
scheme in molecular dynamics \cite{osenda02}.  
The only difference is that here we have to deal with 
the tripolarity of the pairwise potential. 
 
The SU(3)/U(1) symmetry of the potentials leads to the possibility of
a rich variety of clusters: from simple colour dipoles and tripoles 
to strings and complex molecule-like cyclic aggregates. The specific 
configuration of each structure and the number of its constituents could be 
found by calculating the minimum of its effective 
potential. In \cite{yershov05} it was observed that the 
properties of these structures resemble those of 
the fundamental fermions. Based on this observation, 
in this paper we shall further explore two simplest cyclic 
structures, which were identified with the electron and its neutrino.
We shall outline the proposed framework in the next section. 
Afterwards, we shall show that within this framework we are bound to assume
the probabilistic behaviour and intrinsic uncertainty of the systems. 
Then, in Section~4, we shall revisit some properties of the cyclic 
structure identified with the electron. Finally, in Section~5, we shall 
estimate the gyromagnetic ratio of this system.

\section{Two-component basic field}

As a starting point, we suggest that the electric and 
colour fields are unified through their source -- a kind of primitive 
particle, with no properties save its mass and charge. That is, we 
assume that such a particle generates a dual (split) equilibrium 
field, $F(\rho)$, with the following components:
\begin{equation}
F_\ominus(\rho) =s \exp(-\rho^{-1})
\hspace{3ex} {\rm and} \hspace{3ex} 
\hspace{0.4cm} F_\oplus(\rho) =-F'_\ominus(\rho),
\label{eq:sfield0}
\end{equation}
one attractive ($\ominus$) and  the other repulsive ($\oplus$), which are
intimately related to each other by their common origin.
Here the signature $s=\pm1$ indicates the sense of the interaction; 
the derivative is taken with respect to the radial coordinate,
$\rho$. For the sake of simplicity, the amplitude and range 
coefficients in (\ref{eq:sfield0}) are set to unity.
We shall denote the energies corresponding to these components 
of the field (integrals over the entire range of $\rho$) 
as $\tilde{m}_\circ$ and $m_\circ$, respectively.   
Obviously, the second quantity, $m_\circ$, is unity, 
while $\tilde{m}_\circ$ diverges, implying that the above fields 
cannot exist in free states because of their infinite energies.
The approximate antisymmetry between $F_\ominus$ and $F_\oplus$
in the vicinity of their origin (see details in \cite{yershov05}) 
implies that, given a pair of primitive particles with complementary colours/charges, 
these particles will combine into an equilibrium configuration 
(colour-dipole $g^0$, $g^+$ or $g^-$), with the average distance $\rho_\circ$ between 
its components such that the fields balance each other: 
$F_\ominus(\rho_\circ)=-F_\oplus(\rho_\circ)$.
This breaks the initial spherical symmetry of the field, as well as yet 
another fundamental symmetry -- that of scale invariance -- 
since there exists a distance $\rho_\circ$ which can be used 
as the basic unit length for this model.
\begin{figure}[htb]
\vspace{-0.5cm}
 \hspace{-0.3cm}
\begin{turn}{-90}\epsfig{figure=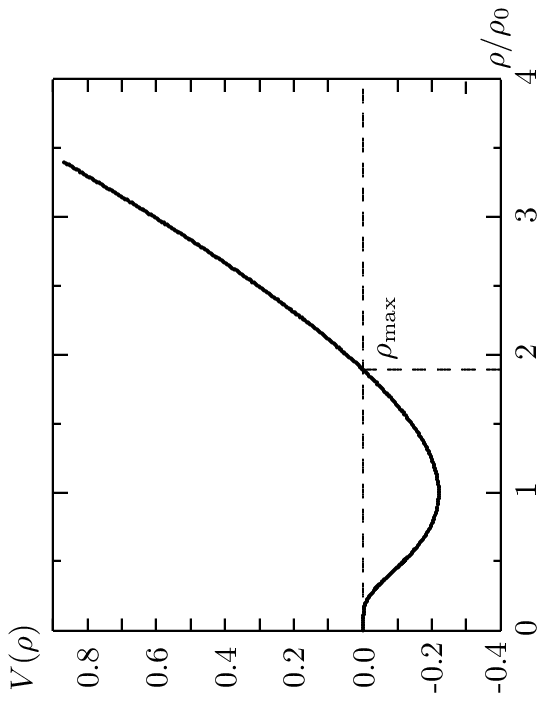, 
 width=6.0cm}\end{turn}
      \hspace{0.3cm}
\begin{turn}{-90}\epsfig{figure=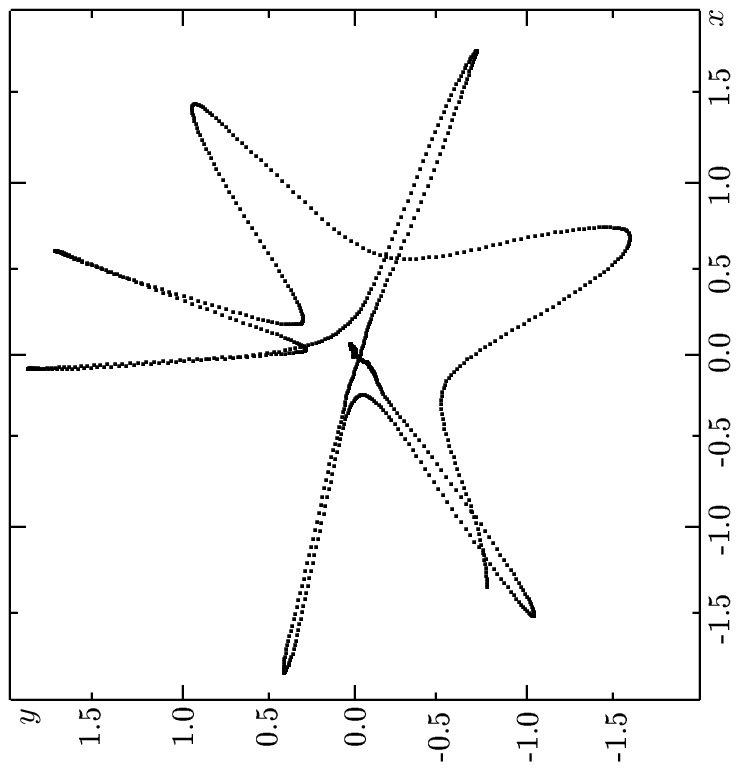,
 width=5.7cm}\end{turn}
\put(-282,-160){\makebox(0,0)[t]{(a)}}
\put(-75,-160){\makebox(0,0)[t]{(b)}}
\put(-199,-108){\makebox(0,0)[t]{\tiny $E_0$}}
\caption{(a): Equilibrium potential $V(\rho)$ based on the field 
(\ref{eq:sfield0}) and used for modelling a two-colour dipole system;\,
(b): Evolution of the two-colour dipole from its initial state
at $\rho=0$ under the stochastic action of an external
system (for $\delta v=0.001v_{\rm max}$). 
The coordinates $x$ and $y$ are measured in units of 
$\rho_\circ$.
 \label{fig:evolution1}
}
\end{figure}
Moreover, it is seen that the potential $V(\rho)$, Fig.\,\ref{fig:evolution1}a,
corresponding to this field
is characterised not only by its inherent length unit,
$\rho_\circ$, but also by the speed and time units, 
$v_\circ$ and $t_\circ$. Indeed,
the speed is calculated as 
\begin{equation}
v(\rho)=\sqrt{\frac{2}{\hat{m}}(E_0-V(\rho))},
\label{eq:speed2particles}
\end{equation}
where $\hat{m}^{-1}=m_1^{-1}+m_2^{-1}$
 is the reduced mass (here $\hat{m}=\frac{1}{2}$ because
for the sake of simplicity in this example we ignore the third colour;
but this does not matter for understanding the point).
The energy $E_0$ of the initial state can be set to zero
(at $\rho=0$), which defines the speed scale for this system 
(unit speed $v_\circ$) through the magnitude of the maximal 
speed, $v_{\rm max}=v(\rho_\circ)\approx 0.937$.
This also establishes the time scale -- by defining the 
unit time, $t_\circ$, such that
$v_\circ t_\circ = \rho_\circ$. Thus, we can see that the 
field (\ref{eq:sfield0}) is fully self-calibrated. Of course, we have to 
take into account the fact that the colour dipole, as with its 
constituents -- colour charges -- cannot exist 
in free states because it has only two of three possible 
diverging components of the field $F_\ominus$ that 
cancel one another.  The colour components could be cancelled
either in a large ensemble of colour dipoles $g^0$ (statistically)
or if three primitive particles were combined together -- 
all with complementary colours and like charges. 
In the latter case, the cancelled energies of three colour-fields 
will be converted into the binding energy of the structure. 
That is, three like-charged primitive particles
will necessarily cohere in a colour-neutral but electrically 
charged singlet -- a $\triangle$-shaped oscillator
\phantom{W}\put(-5,10){\makebox(0,0)[t]{\epsfig{figure=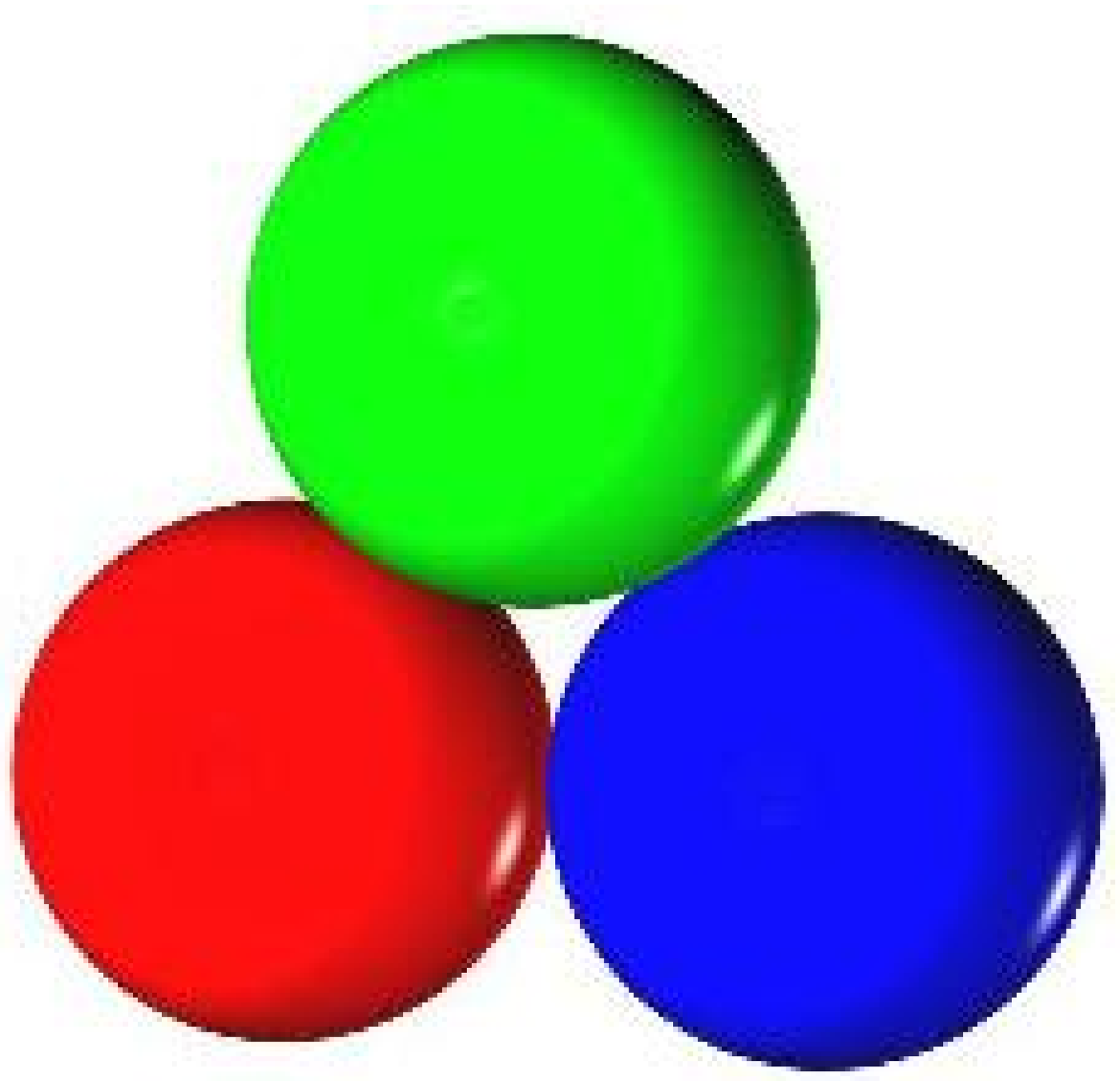,width=0.5cm}}}~
(tripole) 
with radius oscillating near the value
$\rho_{\raisebox{0.1ex}{\tiny $\triangle$}}=
{\rho_\circ}/\sqrt{3}$.
It is noted that the tripole is colour-neutral only at infinity:
nearby its field is colour-polarised since the centres of its 
components do not coincide. This implies that different tripoles 
can be further combined because of their residual chromaticism. 
Using the fields (\ref{eq:sfield0}) one can compute 
the pairwise forces between the colour-charges in a particular 
tripole cluster and estimate the energy of this cluster in order 
to find its equilibrium configuration with minimal energy.  
These calculations can easily be made for some simple configurations 
but the computational difficulties grow dramatically 
with the complexity of the structures. 

\section{Uncertainty of the field}

One can see that the field (\ref{eq:sfield0}) necessarily implies 
an uncertainty, precluding the exact determination of particle 
trajectories. 
Indeed, let us consider the initial state of the simplest system
formed of two primitive charges -- a colour-dipole $g^+$. 
For the sake of simplicity let us ignore for a moment the third 
colour and analyse a two-body problem, which is known to have 
an exact analytical solution. It follows from (\ref{eq:speed2particles}) 
that the dipole components are confined 
within the region $\rho \in (0, \rho_{\rm max})$, Fig.\,\ref{fig:evolution1}a,
where $\rho_{\rm max} \approx 1.894 \rho_\circ$, the particle speed 
vanishing at the ends of this interval. 
As we have already seen, the individual sources of the field (\ref{eq:sfield0}) 
cannot exist in free states because this would lead to their 
infinite energies. This means that we cannot choose 
$\rho=\infty$\, to be the initial state of our system. 
Therefore, the only available natural initial states of the dipole 
correspond to $\rho=\rho_{\rm max}$ and $\rho=0$ (a superposition of particles in the 
origin) with $E_0=0$. The corresponding oscillatory period 
\begin{equation}
T=2(t(\rho_{\rm max})-t(0))=2\sqrt{2\hat{m}}\int\limits_0^{\rho_{\rm max}}
\frac{d\rho}{\sqrt{-V(\rho)}}
\end{equation}
will be infinite, which is what one would expect 
because of a stationary point at the origin,
$F'_\ominus(0)=F'_\oplus(0)=0$. 
This is the bifurcation point of a typical double-well
potential $V(\rho)$, Fig.\,\ref{fig:evolution1}a, 
which is known to lead to chaotic oscillations.
In order to evolve from this state the system requires 
an external (albeit infinitesimally small) action.
Thus, we ought to conclude that this system cannot, in principle,
be treated as isolated. That is, we have to take into account the fact that 
at the initial moment of time an external system (e.g., the rest 
of the universe) adds to (or removes from) our system some infinitesimally 
small portions of energy (which corresponds to the noise with zero expectation 
value). Under this external action, the radius of our system will 
grow from zero to some value, say,  $\delta\rho$, whilst the particle speed 
will increase by the value $\delta v$.  

Within an $\varepsilon$-neighbourhood of the origin 
the proper interactions between the two particles 
will be small compared to the external 
action, so that for some period of time the system will be 
evolving chaotically. After exiting the $\varepsilon$-region, 
this evolution will become more regular, and
the particles will acquire both radial and tangential components 
of their velocities (that is, the initial energy of the system 
will be shared between its angular momentum and oscillatory 
motion).
Outside the $\varepsilon$-region the influence of the external system
will be almost unnoticeable, unless the particles 
occasionally penetrate back into this region. An example of 
the trajectory of a colour-dipole influenced by an external system and
evolving from its initial state with $\rho=0$ is shown in 
Fig.\,\ref{fig:evolution1}b (for $\delta v = 0.001 v_{\rm max}$).
The same reasoning is applicable to the tripolar oscillators whose
evolution will be more complicated, but with the same net result:
it will be impossible to dispose of the influence of the external
system. Thus, we have to arrive at the conclusion that within 
our framework no isolated systems could exist in principle. 
Due to this, even in the simplest case of a two-body system, 
the equations of motion are not analytically solvable, let alone 
the multi-body systems corresponding to more complex 
composite particles. However, given the smallness of the 
perturbations caused by the external system, its influence 
should be noticeable only under extreme conditions when
all the particles are squeezed into a very small volume. 
In most of the other cases, the presence of the external
system will be perceivable only by small deviations from the 
particles' trajectories.

\section{Strings and loops of $\triangle$-shaped tripoles}

\begin{figure*}[htb]
\centering
\hspace{-0.5cm}
\epsfig{figure=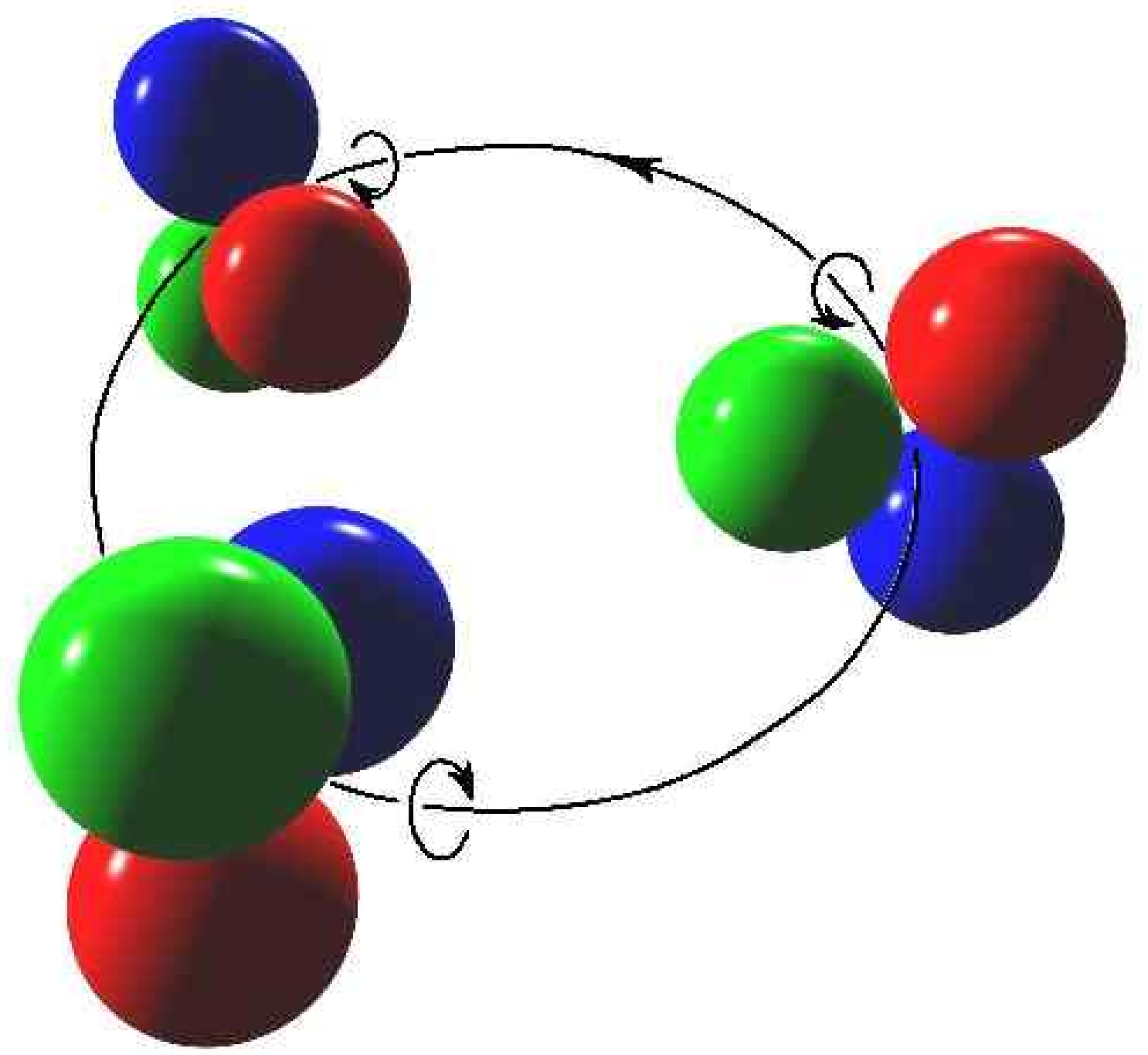,width=3.0cm}
\hspace{0.1cm}
\epsfig{figure=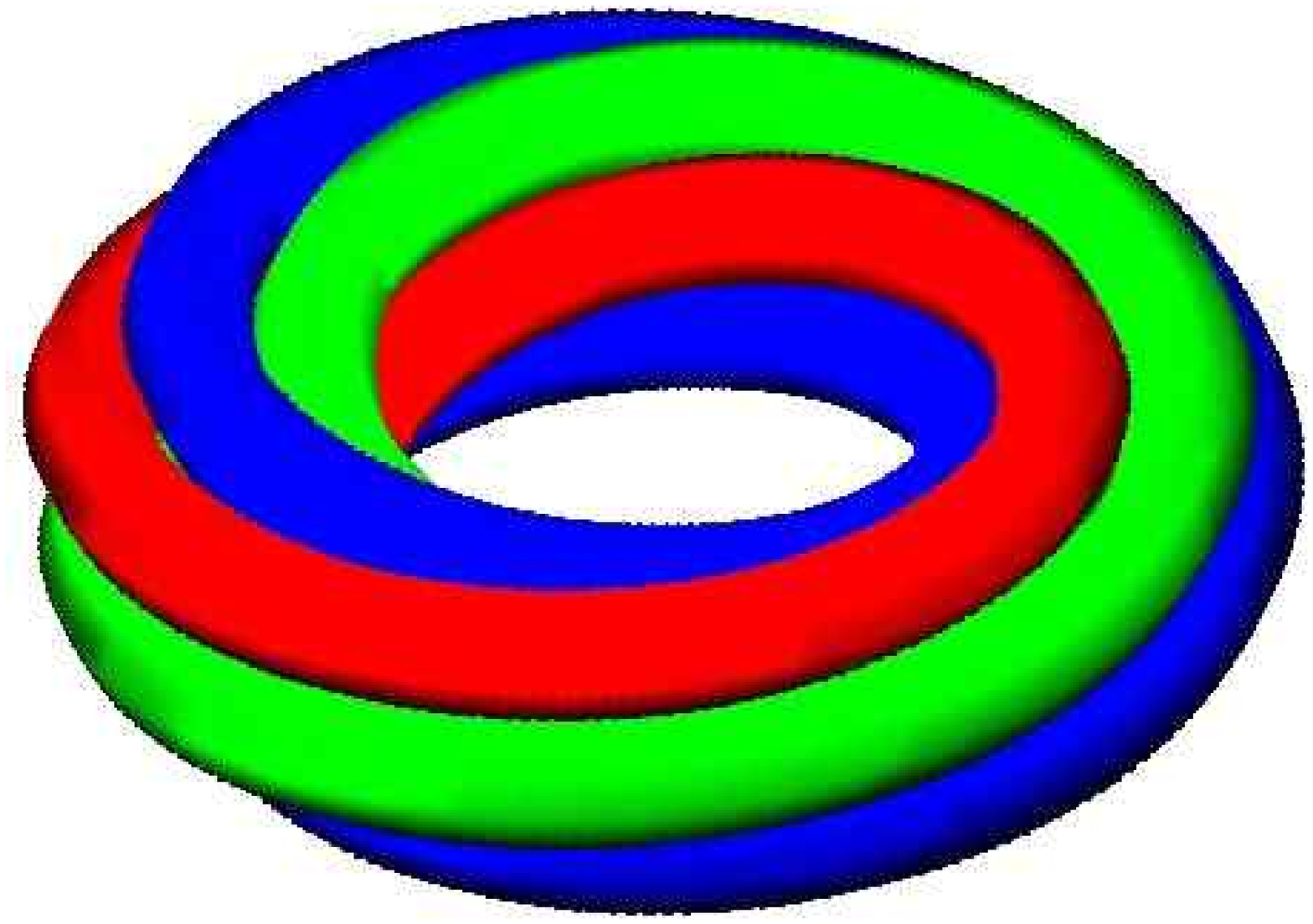,width=3.0cm}
\hspace{0.8cm}
\epsfig{figure=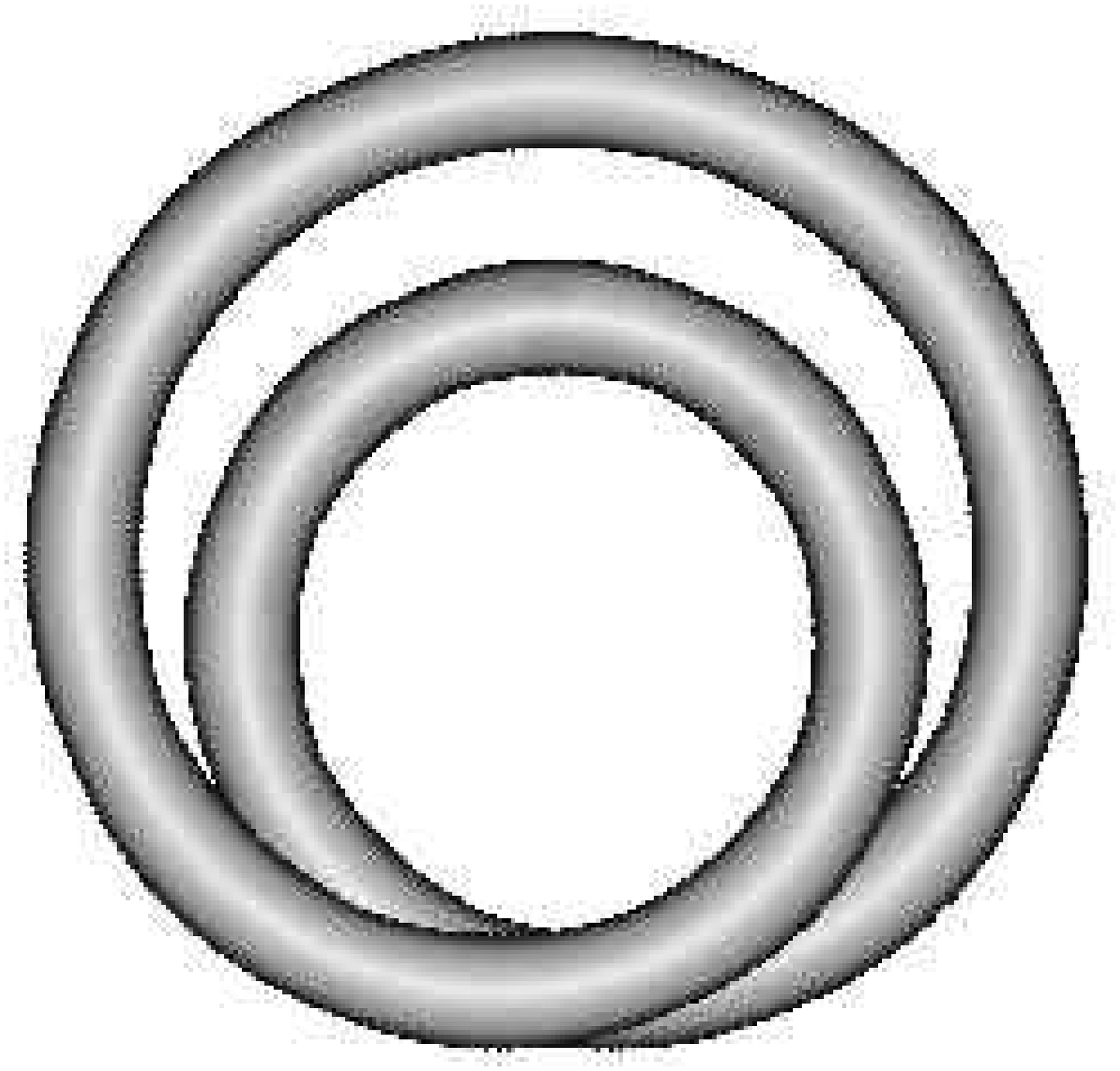,width=2.6cm}
\put(-220,17){\makebox(0,0)[t]{\small (a)}}
\put(-100,17){\makebox(0,0)[t]{\small (b)}}
\put(10,17){\makebox(0,0)[t]{\small {(c)}}}
\caption{(a): Scheme of the triplet ${\tt Y}$ (a chain of three like-charged 
$\triangle$-tripoles closed into a loop); 
(b): trajectories of colour-charges (currents) in the triplet loop; 
(c): trajectory of a single charge.
\label{fig:tripletscheme}
}
\end{figure*}

It was shown in \cite{yershov05} that two $\triangle$-shaped tripoles 
can combine pole-to-pole with each other and form a two-component
oscillator (doublet {\sf d}). 
The sign of the force between the paired tripoles depends 
on their position angle, $\zeta$, with respect to
each other. For example, for $\zeta=\pi/2$ 
the force is vanishing; it is attractive for 
$\pi/2 < \zeta < 3\pi/2$:
\begin{equation}
\put(-110,20){\makebox(0,0)[t]{\epsfig{figure=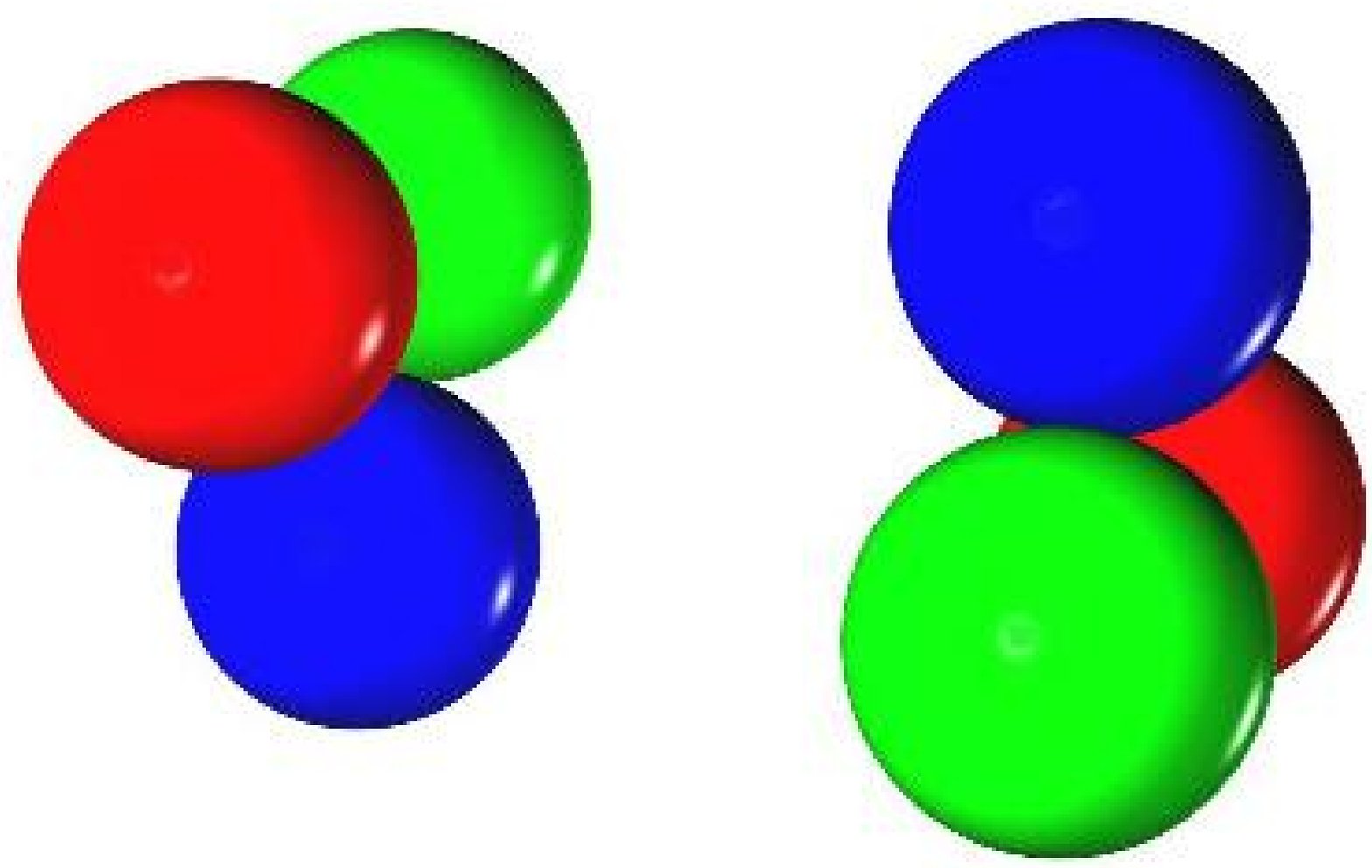,width=1.7cm}}}
\put(-20,20){\makebox(0,0)[t]{\epsfig{figure=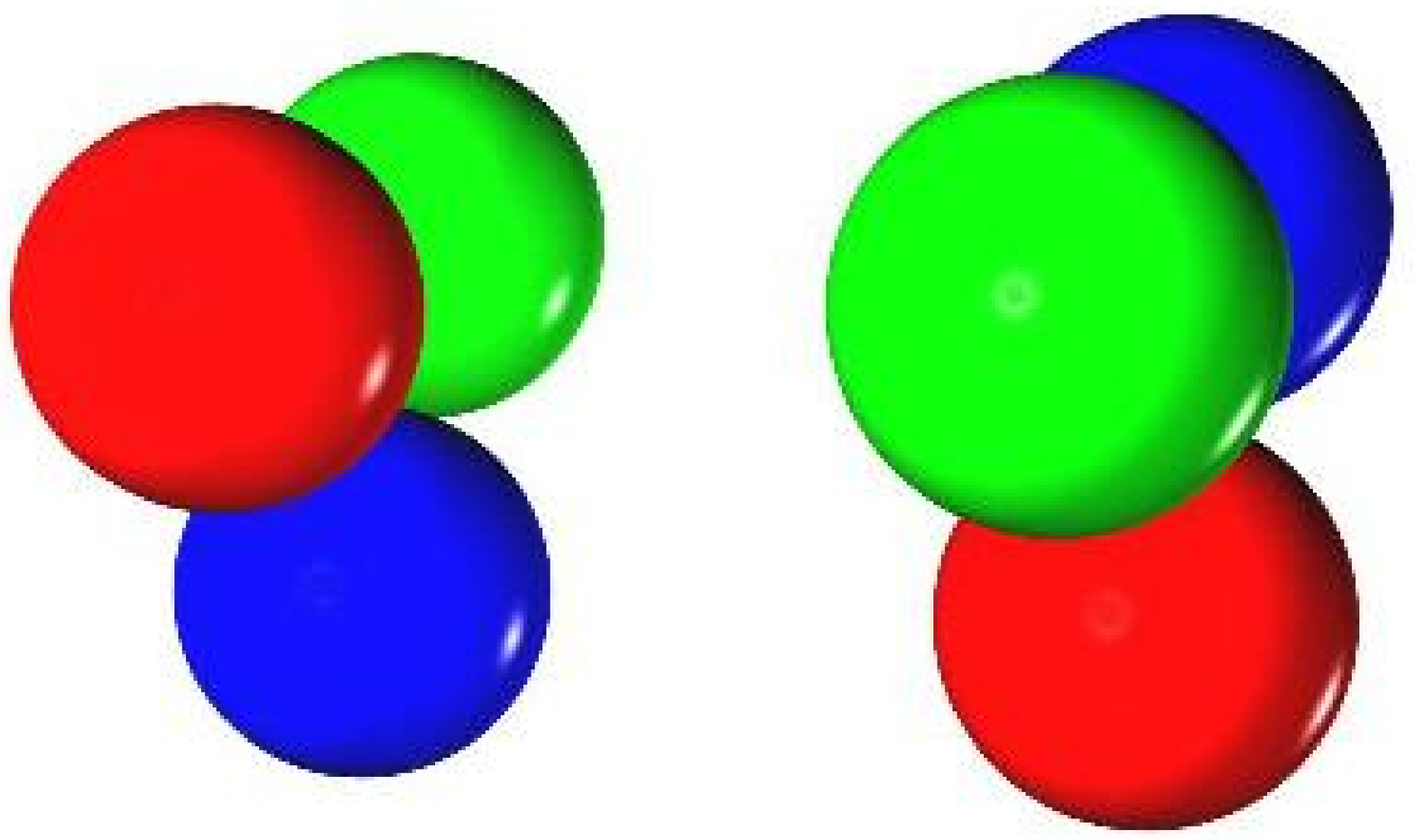,width=1.7cm}}}
\put(70,20){\makebox(0,0)[t]{\epsfig{figure=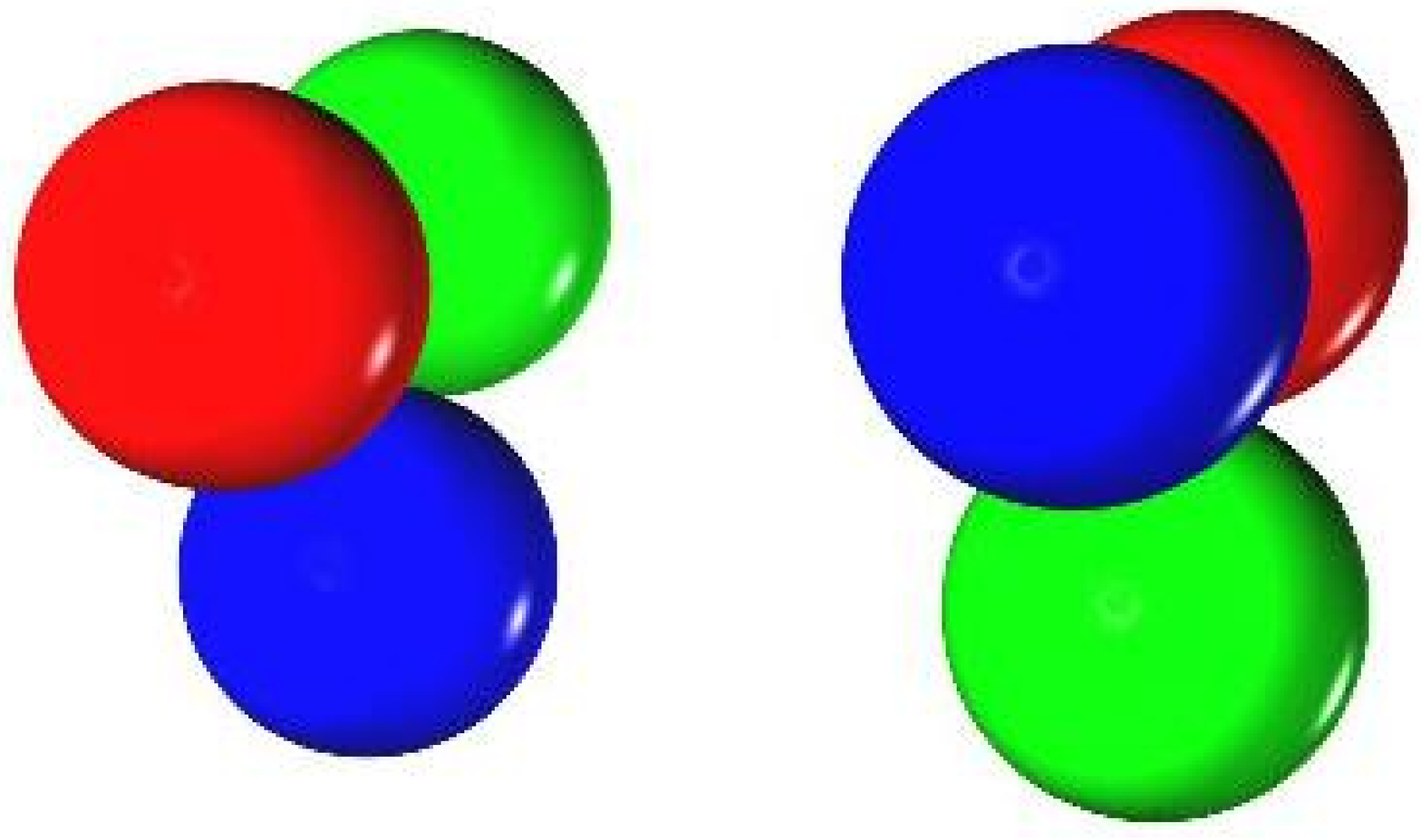,width=1.7cm}}}
\put(-110,20){\makebox(0,0){\curve(12,0, 12,10)}}
\put(-20,20){\makebox(0,0){\curve(12,0, 12,10)}}
\put(70,20){\makebox(0,0){\curve(12,0, 12,10)}}
\put(-75,20){\makebox(0,0){\small $\zeta=\pi$}}
\put(-20,20){\makebox(0,0){\curve(7,-14, 0,-5)}}
\put(-20,20){\makebox(0,0){\curve(12,7,9,6.5,3,4)
\curve(3,4, 4.4,5, 6,6.5)
\curve(3,4, 4.6,4.2, 6.5,4.5)
}}
\put(2,25){\makebox(0,0){$\frac{2\pi}{3}$}}
\put(70,20){\makebox(0,0){\curve(19,-1, 24,7)}}
\put(70,20){\makebox(0,0){\curve(12,7, 17,8, 22,6)
\curve(22,6, 21,6.5, 19,8.5)
\curve(22,6, 21,6.2, 19,6.5)
}}
\put(105,20){\makebox(0,0){$\frac{4\pi}{3}$}}
\put(-110,-17){\makebox(0,0){$\rightarrow \hspace{0.6em} \leftarrow$}}
\put(-20,-17){\makebox(0,0){$\rightarrow \hspace{0.6em} \leftarrow$}}
\put(70,-17){\makebox(0,0){$\rightarrow \hspace{0.6em} \leftarrow$}}
\put(-150,6){\makebox(0,0){${\sf d}^+=$}}
\put(-60,2){\makebox(0,0){${\sf d}^+_\down=$}}
\put(30,2){\makebox(0,0){${\sf d}^+_\up=$}}
\vspace{1.2ex}
\label{eq:dipoleattractive}
\end{equation}
%
 and repulsive for $|{\zeta}|<\pi/2$:
\vspace{1.5ex}
%
\begin{equation}
\put(-110,10){\makebox(0,0)[t]{\epsfig{figure=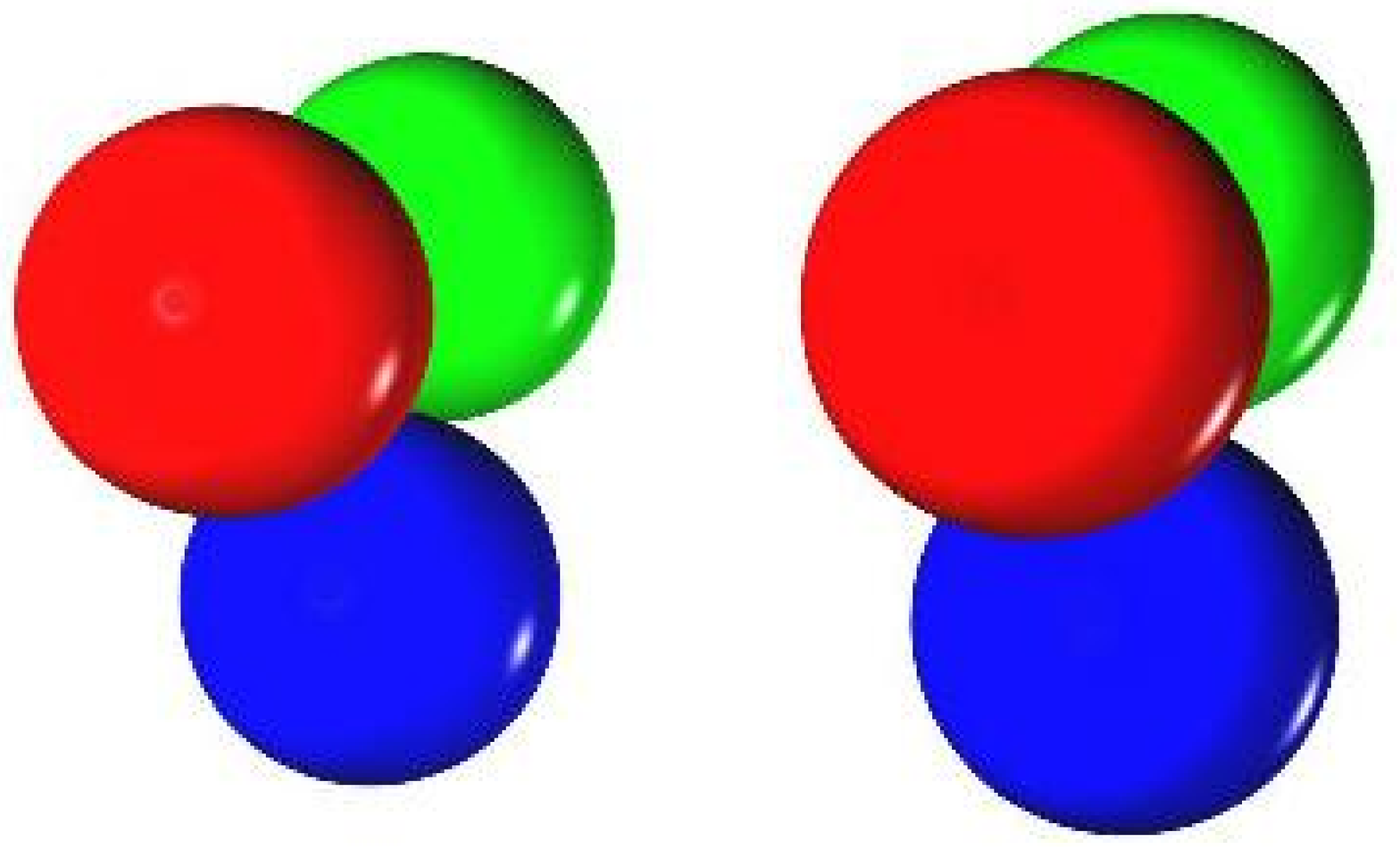,width=1.7cm}}}
\put(-20,10){\makebox(0,0)[t]{\epsfig{figure=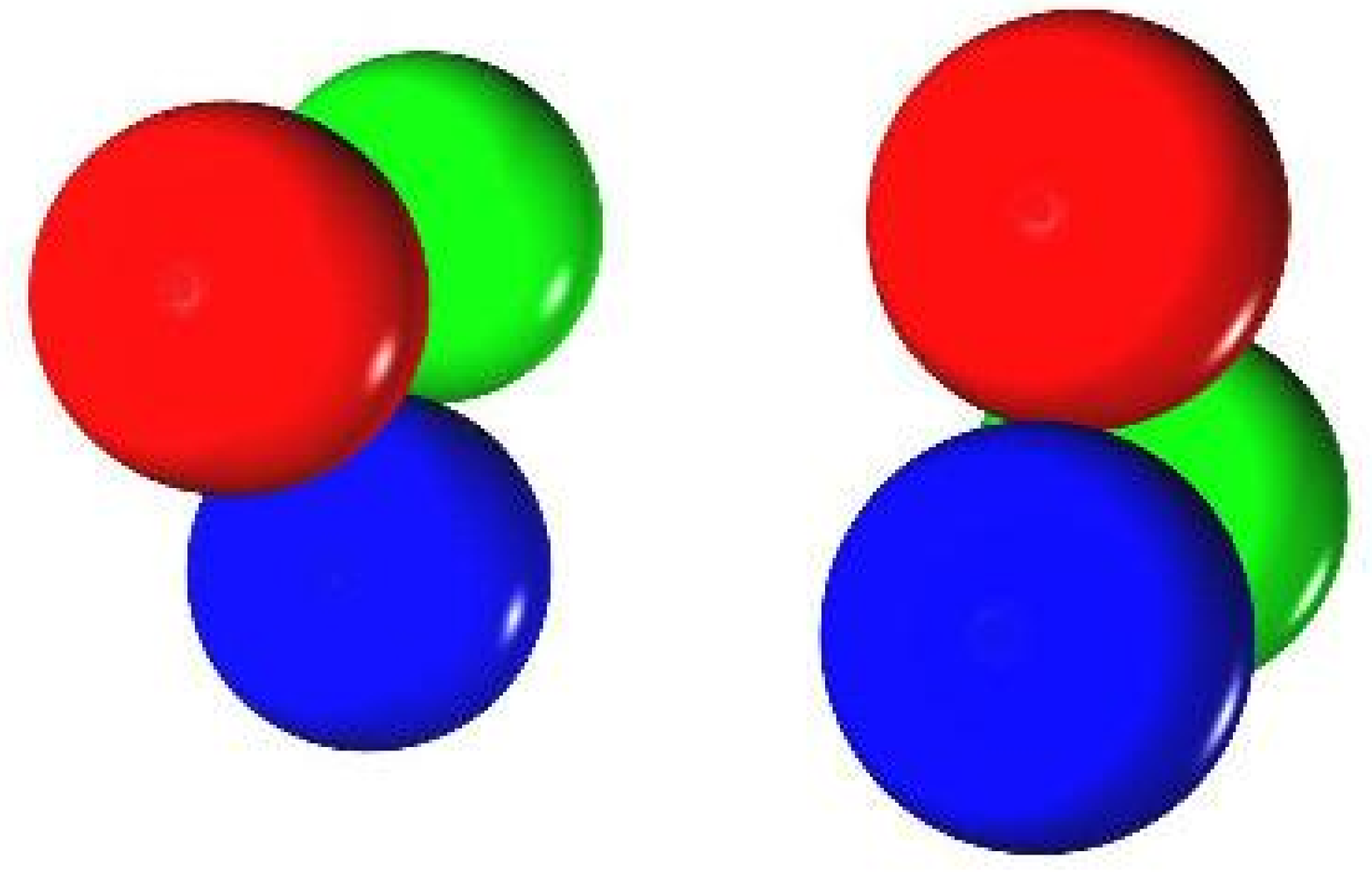,width=1.7cm}}}
\put(70,10){\makebox(0,0)[t]{\epsfig{figure=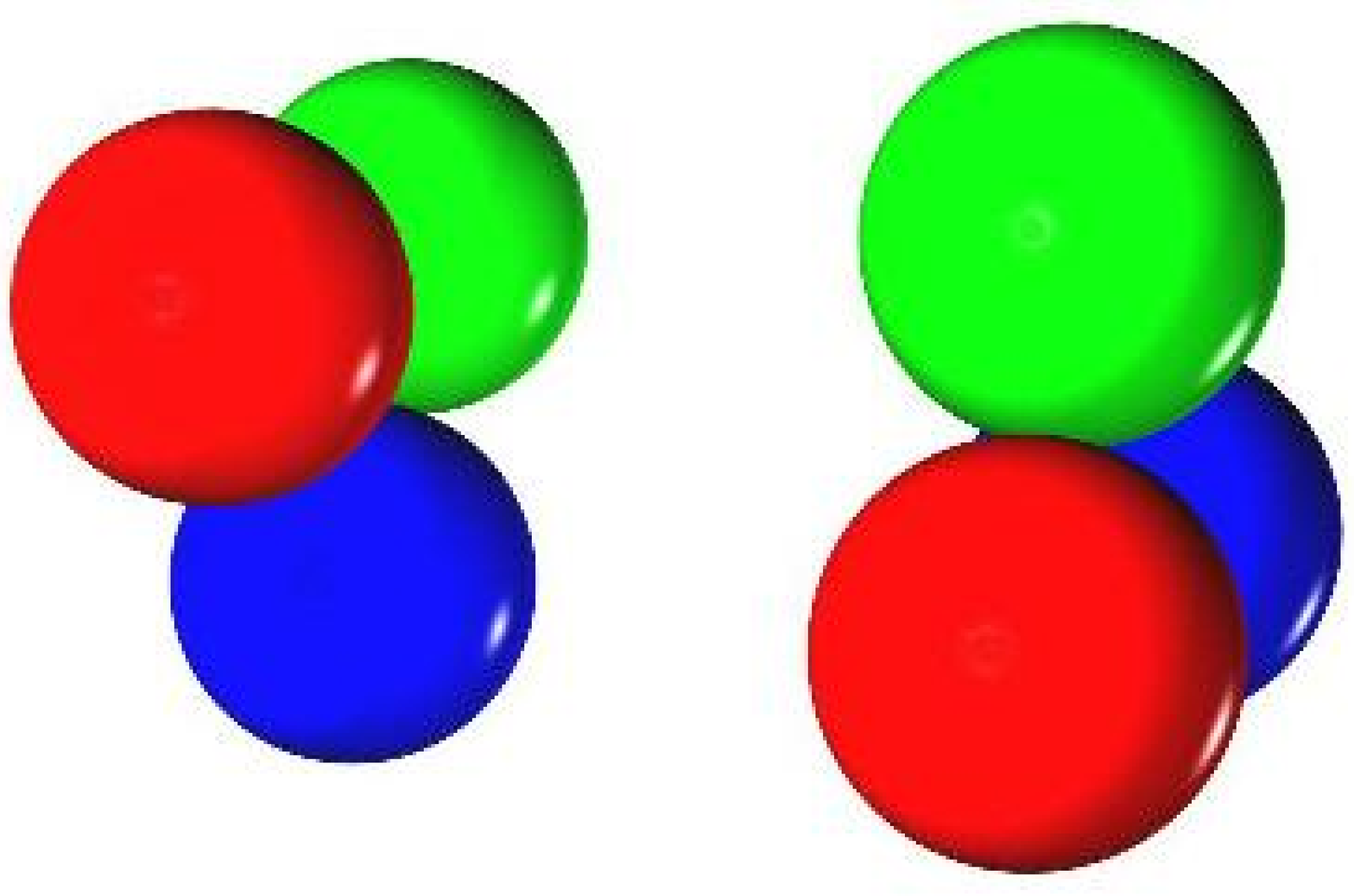,width=1.7cm}}}
\put(-108,-20){\makebox(0,0){\curve(11,-8, 11,-20)}}
\put(-18,-20){\makebox(0,0){\curve(12,-9, 12,-19)}}
\put(72,-20){\makebox(0,0){\curve(12,-9, 12,-19)}}
\put(-80,-27){\makebox(0,0){\small $\zeta=0$}}
\put(-10,-9){\makebox(0,0){\curve(-1,-21, -6,-30)}}
\put(-18,-20){\makebox(0,0){\curve(12,-16, 6,-16, 3,-12)}}
\put(-18,-20){\makebox(0,0){\curve(2.3,-10.8, 2.8,-12.8, 3.3,-14.5)}}
\put(-18,-20){\makebox(0,0){\curve(3.0,-8.7, 3.9,-9.8, 5.8,-10.9)}}
\put(-28,-28){\makebox(0,0){$-\frac{\pi}{3}$}}
\put(90,-10){\makebox(0,0){\curve(-1,-12, 3,-18)}}
\put(72,-20){\makebox(0,0){\curve(12,-14, 16,-12, 19.5,-4.5)}}
\put(72,-20){\makebox(0,0){\curve(19.4,2.5, 19.4,1.5, 19.3,-1.2)}}
\put(72,-20){\makebox(0,0){\curve(19.4,2.5, 18.7,1.5, 16.9,-0.5)}}
\put(93,-32){\makebox(0,0){$\frac{\pi}{3}$}}
\put(-113,10){\makebox(0,0){$\leftarrow \hspace{2.7em} \rightarrow$}}
\put(-21,10){\makebox(0,0){$\leftarrow \hspace{2.7em} \rightarrow$}}
\put(70,10){\makebox(0,0){$\leftarrow \hspace{2.7em} \rightarrow$}}
\vspace{1.5ex}
\label{eq:dipolerepulsive}
\end{equation}
%
Thus, separated by distance  $\rho >2\rho_\circ$, 
two tripoles will tend to combine into a doublet configuration 
(${\sf d}^+$ or ${\sf d}^-$) with $\zeta=\pi$. 
The neutral doublet
\begin{equation}
\put(-20,20){\makebox(0,0)[t]{\epsfig{figure=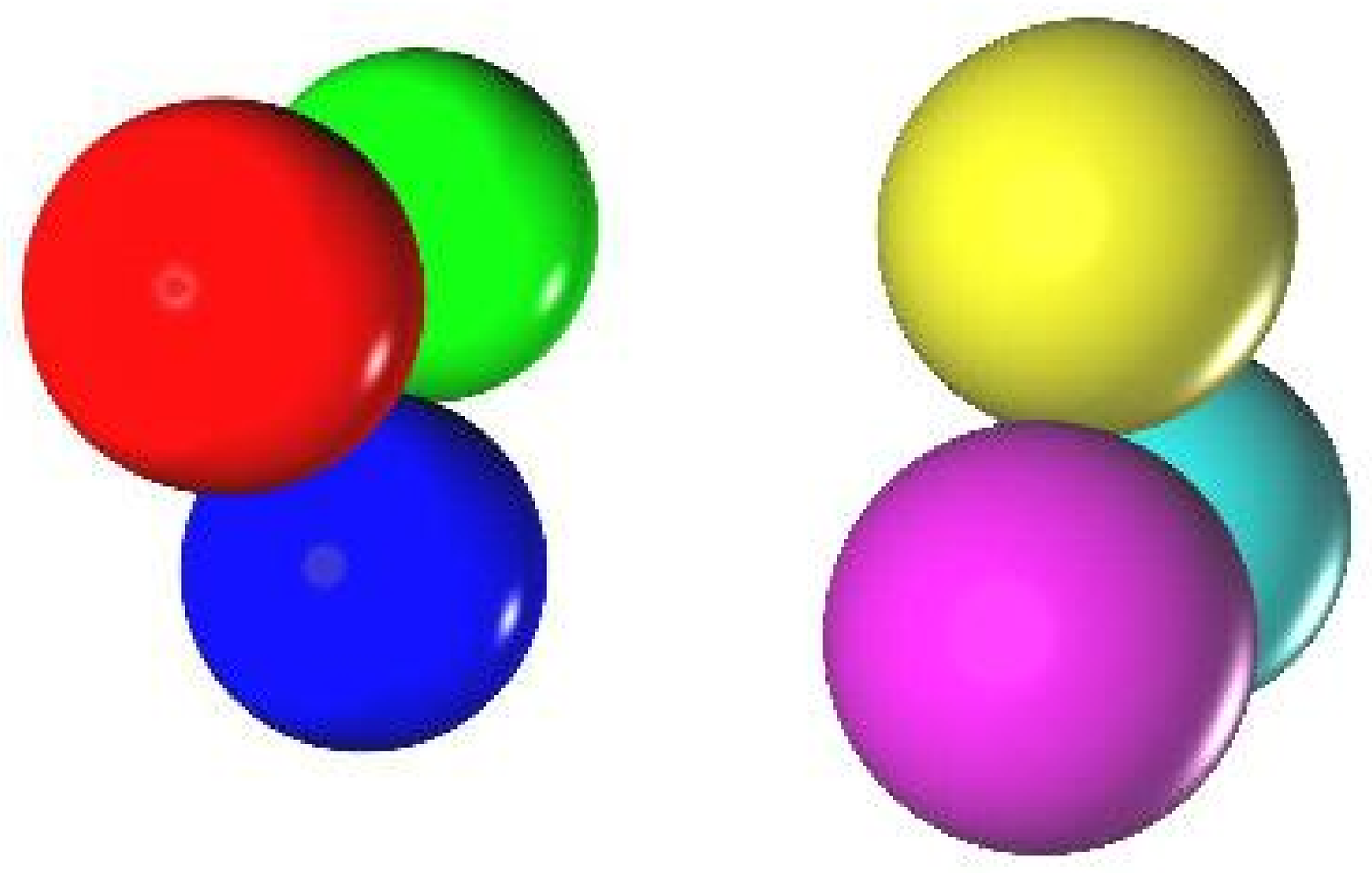,width=1.7cm}}}
\put(-60,6){\makebox(0,0){${\sf d}^\circ=$}}
\put(-38,20){\makebox(0,0){\small $+$}}
\put(2,20){\makebox(0,0){\small $-$}}
\label{eq:deltazero}  
\end{equation}
can also be formed. 
The width of the potential well for $\zeta=\pi$ and 
$\rho > 2\rho_\circ$ allows a certain degree of rotational freedom 
for the paired tripoles, so that the position angle can oscillate 
within $\frac{2}{3}\pi < \zeta < \frac{4}{3}\pi$.
We shall use the symbols $\up$ and $\down$ to denote, respectively, 
the clockwise and anticlockwise directions of rotation.
Then, the rotational oscillations of the doublet can be 
represented as
\begin{equation}
{\sf d}^+_\up  \hspace{0.2cm} 
\rightleftarrows
\hspace{0.2cm} 
{\sf d}^+_\down 
\hspace{0.5cm} {\rm or} \hspace{0.5cm}
{\sf d}^\circ_\up  \hspace{0.2cm} 
\rightleftarrows
\hspace{0.2cm} 
{\sf d}^\circ_\down ~.
\label{eq:deltaprotation}  
\end{equation}
The $\zeta$-dependence of both 
strength and sign of the bond force between the tripoles 
implies that the distance 
$\rho$ is covariant with the position angle $\zeta$,
i.e., that the translational and rotational oscillations of 
the doublet {\sf d} are synchronous. 
It follows that due to the $\frac{2\pi}{3}$-symmetry of the tripoles
their rotations in a chain of like-charged tripoles can cyclically repeat
after each three links: 
\phantom{WW}\put(-7,10){\makebox(0,0)[t]{\epsfig{figure=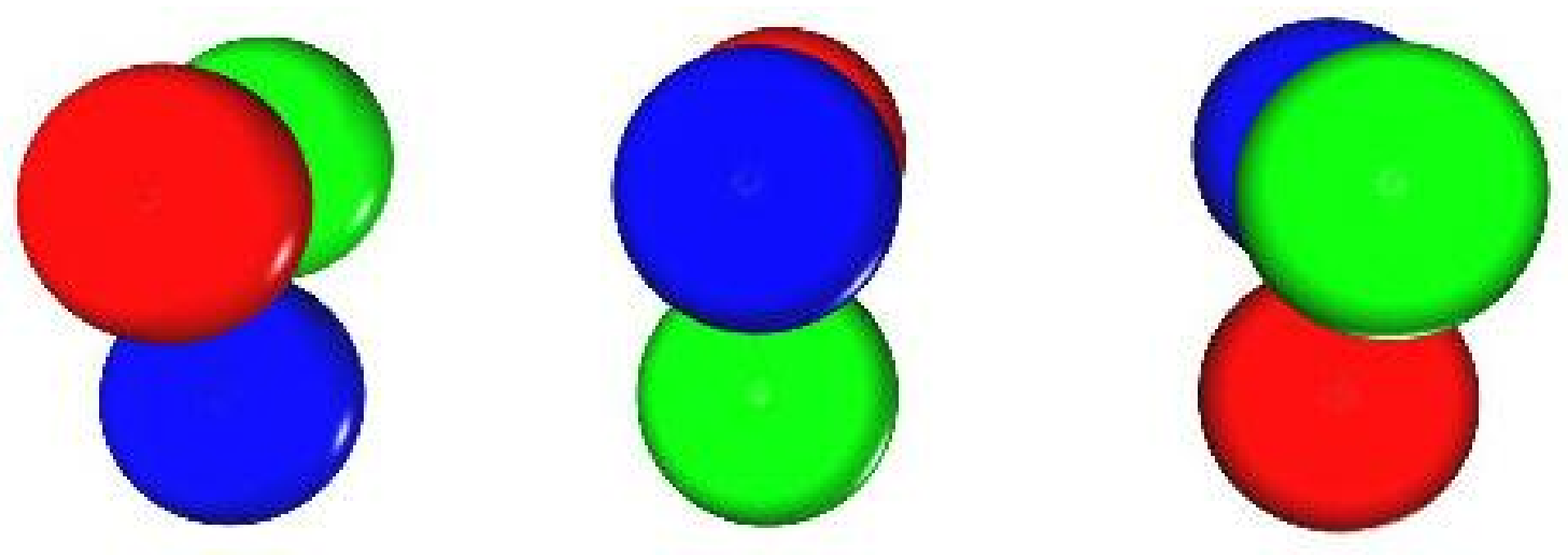,width=1.2cm}}}~~~,
leading to the closure of the chain in a symmetric loop (triplet), 
Fig.\,\ref{fig:tripletscheme}a,
denoted here as ${\tt Y}= 3 \triangle$\hspace{0.3ex}.
The consecutive $\frac{2\pi}{3}$-phase-shifts of the 
tripoles in this chain can be either clockwise 
or anticlockwise corresponding to two 
possible helical states of the triplet, ${\tt Y}_\up$ and ${\tt Y}_\down$.

\begin{figure}[htb]
\centering
\epsfig{figure=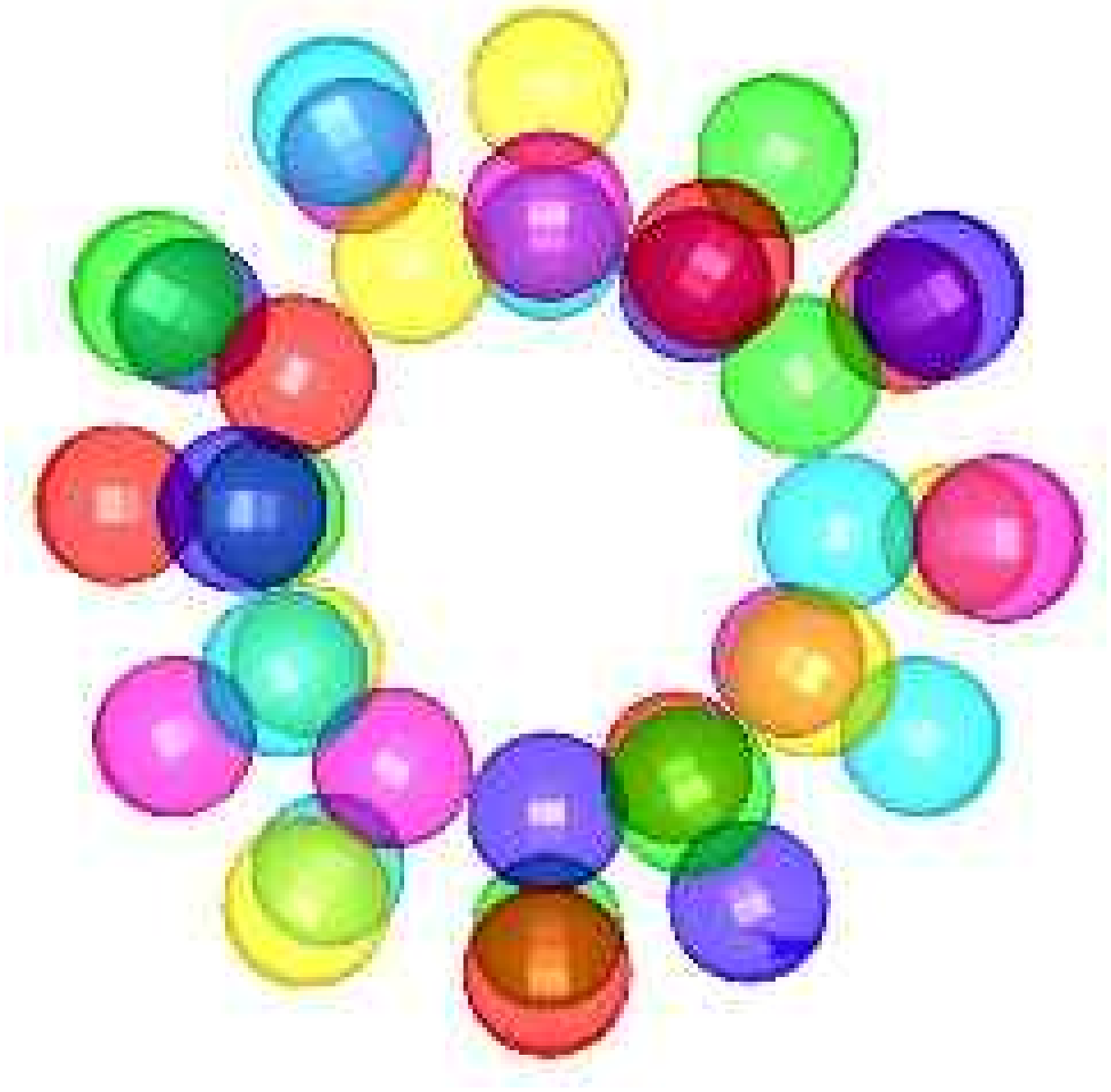,width=3.cm}%
\epsfig{figure=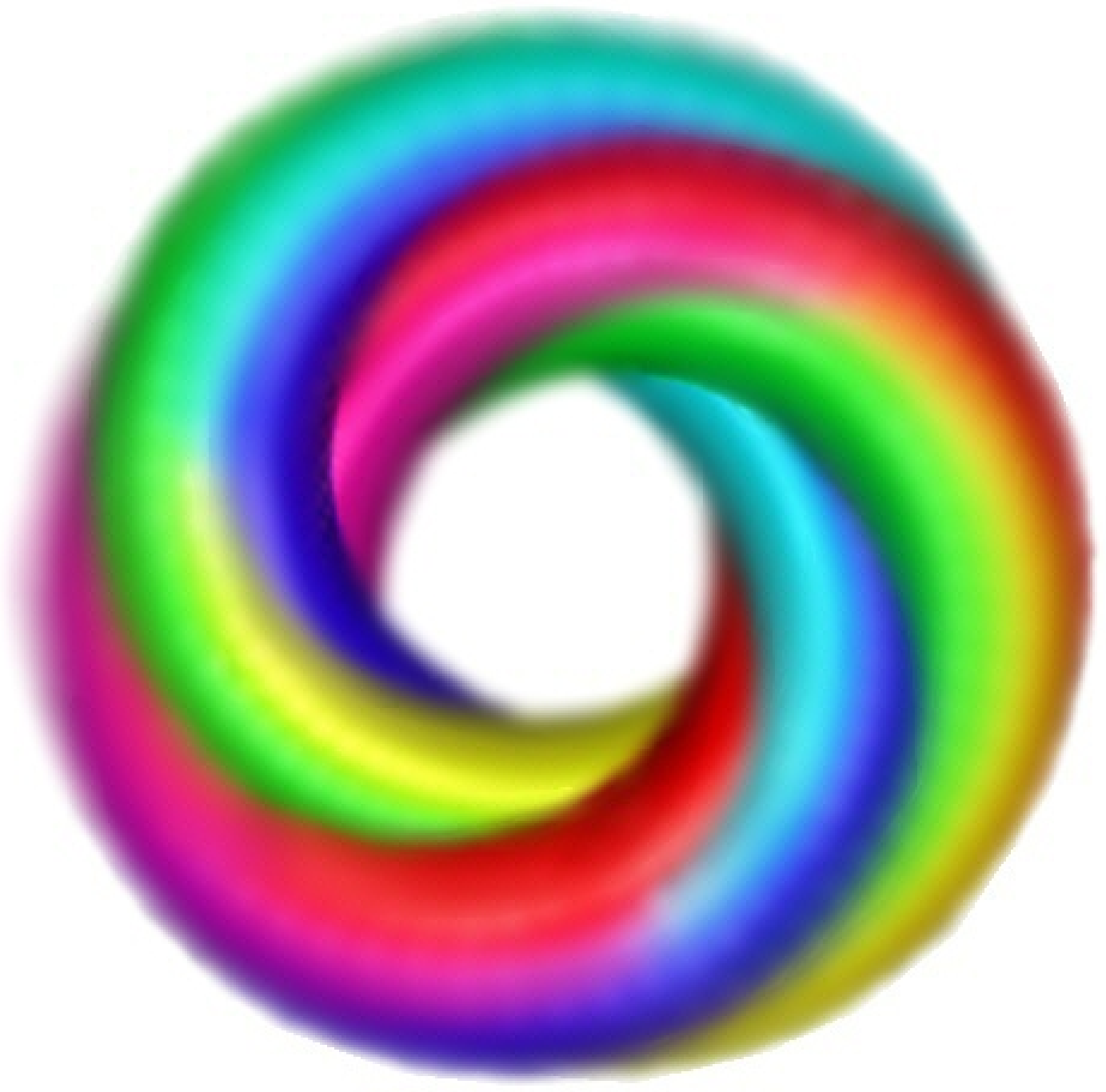,width=4.5cm}%
\put(-120,10){\makebox(0,0)[t]{\small (a)}}
\put(10,10){\makebox(0,0)[t]{\small (b)}}
\caption{
(a):\, Scheme of the hexaplet ${\tt X}$ -- a loop configuration of six
tripole-antitripole pairs; and (b):\,
Trajectories of colour-charges (currents) in this structure.
The antitripoles are coded with lighter colours.
 \label{fig:eneutrino}
} 
\end{figure}

In a similar way, one can find that, given a
chain of unlike-charged pairs of tripoles, the
pattern of rotations in this chain repeats after each six 
tripole-pairs, leading to the closure of the chain
in a six-component loop (hexaplet), Fig.\,\ref{fig:eneutrino}a,
which we shall denote as ${\tt X} =6{~\DP~\DN}$\hspace{0.3ex}. 
Obviously, this structure is electrically neutral 
and, like the triplet, can also be found in one of 
two possible helical states, ${\tt X}_\up$ or ${\tt X}_\down$. 

It is seen that the charges constituting a cyclic 
structure can spin around its ring-closed axis. 
In the case of the triplet, 
${\tt Y}$, this will generate a toroidal (ring-closed) 
magnetic field which will force these 
charges to move along the torus.
This orbital motion will generate a secondary (poloidal)
magnetic field, contributing to the spin of these charges 
around the ring-axis, and so forth, until the charges  
reach their maximal speed, say, $v_\circ$. Such a dynamo mechanism 
for generating a self-consistent magnetic field is
studied in detail in astrophysics \cite{parker78} and solar physics
\cite{pipin00}. It is also used for stabilisation of toroidal 
plasma flows in the tokamak fusion reactors 
\cite{kadomtsev93}.
The only difference between the 
standard dynamo models and our case is that 
the ${\tt Y}$-structure here does not require any 
external angular momentum to maintain its magnetic field.       
The trajectories of charges (currents) 
are clockwise, ${\tt Y}_\up$, or anticlockwise,  
${\tt Y}_\down$, helices (Smale-Williams curves), 
which, by their closure,  make a $\pi$-twist around the 
ring-closed axis of the structure (Fig.\,\ref{fig:tripletscheme}b). 
Such a twisting dislocation of the phase is known as the topological 
charge \cite{kleman83}, also called the 
dislocation index, which has a sign corresponding 
to the winding direction (clockwise 
or anticlockwise) and the magnitude related 
to the winding number per $2\pi$-orbit path.
In these terms, the $\pi$-phase shift of the currents in the 
structures ${\tt X}$ and ${\tt Y}$ corresponds to the
topological charge $S=\pm \frac{1}{2}$.

It is worth noticing that, since the $\frac{2}{3}\pi$-symmetry
of the tripole is reproduced on a higher hierarchical level
-- in the structure ${\tt Y}$ --  
the path of each colour-charge belonging to a particular tripole
overlaps exactly with the paths of two other colour charges that
belong to two other tripoles and whose colours are complementary
to the colour of the first charge. This means that the trajectories 
of charges (currents) in the structure ${\tt Y}$ are dynamically
colourless (Fig.\,\ref{fig:tripletscheme}c). 
That is, averaged in time, the field of the triplet ${\tt Y}$
would have only two (positive and negative) polarities corresponding to 
the conventional electric field. 
The same symmetry is also observed in the hexaplet 
whose currents are shown in Fig.\,\ref{fig:eneutrino}b.
There are twelve current loops in this structure, six negative 
and six positive, compared to the three unipolar loops in the triplet ${\tt Y}$.

The colourlessness of the time-averaged fields of ${\tt Y}$ and ${\tt X}$
does not necessarily imply that these particles cannot colour-interact
with each other. On the contrary, if the motions of their constituents 
are synchronised, these structures will induce 
an attractive or repulsive force towards each other, additional to the 
conventional electrostatic force.
Thus, given a pair of triplets ${\tt Y}$
(or hexaplets ${\tt X}$) with opposite helicities (${\tt Y}_\up {\tt Y}_\down$),
the mutual orientation of the tripoles in the 
pair corresponds to an attractive force between them
[see the diagram (\ref{eq:dipoleattractive})], whereas like-helicities
(${\tt Y}_\up {\tt Y}_\up$ or ${\tt Y}_\down {\tt Y}_\down$) 
correspond to repulsion [diagram (\ref{eq:dipolerepulsive})].
By contrast, one can find \cite{yershov05} that the pattern of repulsion and 
attraction between the constituents is reversed in a mixed pair 
(${\tt X}$ with ${\tt Y}$), in which like-topological charges 
attract each other and unlike ones repel. As a result, the combination 
${\tt X}_\up{\tt Y}_\up$ (or ${\tt X}_\down{\tt Y}_\down$) 
has an integral topological charge ($S_{\tt XY}=\pm 1$). 
The combination ${\tt X}_\up{\tt Y}_\down$ would have a zero topological
charge, but this system is unlikely to exist, since the 
topological charges of ${\tt X}_\up$ and 
${\tt Y}_\down$ in this system are repulsive to each other.
The clustering of ${\tt Y}$-particles (in our interpretation -- electrons)
is, in fact, observed experimentally: it is known that the electrons can
form clusters, e.g, in extended media where they may undergo crystallisation 
at low densities, as has been predicted by E.\,Wigner \cite{wigner34} and 
then shown by different research groups \cite{maksym96}.  

It is interesting to note that the momentum and angular
momentum of the hexaplet ${\tt X}$
in the structure ${\tt XY}$ are coupled to each other through the magnetic 
field (see \cite{yershov05} for details). Given also the possibility of 
polarisation of the hexaplet ${\tt X}$ when it is combined with the 
triplet ${\tt Y}$, the above coupling of momenta would result in 
the conjugation of charge and parity of the particles ${\tt X}$ and ${\tt Y}$ 
(at the moment when they leave the system ${\tt XY}$). 
Such a conjugation, known as CP-symmetry, is observed in the $\beta$-decay products, 
$e^-$ and $\overline{\nu}_e$. Since we have already identified the structures 
${\tt Y}$ and ${\tt X}$ with, respectively, $e$ and $\nu_e$, we can see 
that our model provides a natural explanation of the CP-symmetry 
and of the neutrino left-handedness. 

It is also found that the repulsive (or attractive) force 
between two helical structures is maximal when their topological
charges have half-integer magnitudes. This force 
diminishes when the magnitude of one (or both) of 
the topological charges deviates from the half-integer value
and eventually decays to zero
when the magnitude of any of the topological charges 
takes an integral value. 
This pattern of attraction and repulsion adheres to the Pauli 
exclusion principle, and here we have deliberately chosen 
the symbols $\up$ and $\down$ to denote the opposite helicities
(topological charges), implying that the helicity of a cyclic 
structure is equivalent to the quantum notion of spin.
This conjecture is also supported by the fact that 
quantum spin is measured in units of angular momentum ($\hbar$),
and so too is the topological charge in question, which is derived 
from the rotational motion of the tripoles $\triangle$ around 
the ring-closed axis of the triplet ${\tt Y}$ or hexaplet ${\tt X}$.

\section{Magnetic moment of the triplet}

\begin{figure}[htb]
\centering
\includegraphics[scale=0.9]{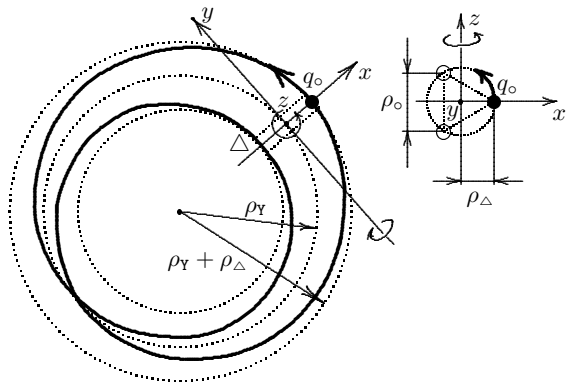}
\caption{Polar view of the trajectory of one of the charges ($q_\circ$)
belonging to a $\triangle$-tripole and moving within the structure ${\tt Y}$. 
The tripole is spinning around its polar axis ($y$)
and precessing around the axis $z$ (perpendicular to the 
orbit plane). 
 \label{fig:oneloop}
} 
\end{figure}

It is instructive to estimate the magnetic moment of the looped currents
of the triplet ${\tt Y}$, together with the corresponding gyromagnetic ratio. 
As we have already noted, the charges moving within this structure 
trace helical trajectories, which, by the closure of 
their $2\pi$-paths along the ring-axis of ${\tt Y}$
are additionally $\pi$-twisted around this axis. 
This means that each of these currents is formed of 
two loops (see Fig.\,\ref{fig:oneloop}).  
\noindent
The magnetic moment created by these loops can be calculated as
\begin{equation*}
\mu_{\tt Y} \approx I_{\rm int} A_{\rm int}+I_{\rm ext} A_{\rm ext},
\end{equation*}
where $I_{\rm int}$ and $I_{\rm ext}$ are the 
currents corresponding to the ``internal'' (smaller) and ``external''
(larger) loops; $A_{\rm int}$ and $A_{\rm ext}$ are the loop 
areas.
We can reasonably approximate the loop radii by the values 
\begin{equation*}
\rho_{\rm int}=\rho_{\tt Y}-\frac{\rho_\triangle}{2} 
\hspace{0.5cm} {\rm and} \hspace{0.5cm} 
\rho_{\rm ext}=\rho_{\tt Y}+\frac{\rho_\triangle}{2}. 
\end{equation*}
Given
\begin{equation*}
I=\frac{3q_\circ v}{2\pi \rho} \hspace{0.5cm} {\rm and} 
\hspace{0.5cm}A=\pi \rho^2,
\end{equation*}
the magnetic moment of the structure will be 
\begin{equation}
\begin{split}
\mu_{\tt Y}&=\frac{3q_\circ v_{\rm int}}{2\pi(\rho_{\tt Y}-\rho_\triangle/2)}
\cdot \pi (\rho_{\tt Y}-\rho_\triangle/2)^2 + \\
&+\frac{3q_\circ v_{\rm ext}}{2\pi(\rho_{\tt Y}+\rho_\triangle/2)}
\cdot \pi (\rho_{\tt Y}+\rho_\triangle/2)^2 = \\
&=\hspace{0.3ex}3q_\circ\omega(\rho_{\tt Y}^2+\frac{\rho_\triangle^2}{4}),
\end{split}
\label{eq:magmoment}
\end{equation}
where $v_{\rm int}=\omega(\rho_{\tt Y}-\rho_\triangle/2)$ 
and $v_{\rm ext}=\omega(\rho_{\tt Y}+\rho_\triangle/2)$ 
are the averaged orbital speeds of the charges; $\omega$ is the angular
speed of the tripoles; and the coefficient ``3'' in the expressions above 
appears due to the 
fact that there are actually three charges (of different colours) 
moving along the same path (but this is not essential here).
The corresponding angular momentum 
\begin{equation}
\ell_{\tt Y} \approx 3m_\circ \rho_{\tt Y} v_\circ=
3m_\circ \omega \rho_{\tt Y}^2
\label{eq:angmomentum}
\end{equation}
results in the following gyromagnetic ratio of the triplet:
\begin{equation}
{\rm g}_{\tt Y}=\frac{2m_\circ}{q_\circ}\cdot\frac{\mu_{\tt Y}}{\ell_{\tt Y}}
\approx 2 \left(1+\frac{\rho_\triangle^2}{4\rho_{\tt Y}^2}\right).
\label{eq:gfactor}
\end{equation}
Since 
${\rho_\triangle^2}/{4\rho_{\tt Y}^2}\ll 1$, 
the gyromagnetic ratio above is approximately
equal to the number of current loops,
 ${\rm g}_{\tt Y} \approx 2$. To derive this quantity more 
accurately one has to account for the detailed solenoidal geometry of the 
currents in the triplet, as well as for the radial oscillations of its
constituents and of the whole structure. 

Of course, the proportionality of the magnetic moment to the loop
number of a solenoid is commonplace. In fact, what we have shown in
Eq.\,(\ref{eq:gfactor}) is that the gyromagnetic ratio of the triplet
is slightly larger than 2, which agrees with the experimental value
\begin{equation}
{\rm g}_e^{\rm exp}\approx 2.002319
\label{eq:geexp}
\end{equation}
of the Land\'e g-factor for the electron. 
The value ${\rm g}_e=2$ was explained quantum-mechanically by P.A.M.Dirac 
\cite{dirac28a}, and, hitherto, it was assumed that this value could 
not be explained in terms of classical mechanics because the 
classically-derived g-factor should be equal to unity. We can see 
now that this is not necessarily so. In fact, the derivation
of the half-integral spin for a system with an integer 
classical orbital momentum was already reported in \cite{kovachev04} a few
years ago. Here we have shown that a classical approach can be 
even closer to reality. From (\ref{eq:gfactor}) and  (\ref{eq:geexp}) 
we can see that $\rho_{\tt Y}/\rho_\triangle\approx 15$, corresponding to 
the shape of a rather thin O-ring.   

\section{Discussion}
Without invoking any {\it ad hoc} assumptions and based
solely on first-principles, i.e., on our conjecture about the symmetry 
of the basic field (\ref{eq:sfield0}), we have found that practically 
all the properties of the emerging triplet structure ${\tt Y}$, 
including its gyromagnetic ratio, match those of the electron, which 
makes it natural to identify the triplet with the electron.
Likewise, the properties of the hexaplet ${\tt X}$ suggest identifying it with 
the electron-neutrino. Incidentally, since the discovery 
of the electron many physicists explored the idea of a spinning structure in the
form of a disk or ring \cite{parson15} in order to 
explain the properties of the electron, which led to important discoveries.
For example, it was A.Compton's intention to determine the size of 
the ring-electron that inspired him to perform his famous 
scattering experiments \cite{compton19}. 
The early models of the ring-electron were created as straightforward 
interpretations of the observed properties of this particle
but they could not explain the magnitude of its magnetic moment.
 
Some modern physicists have revisited the old ideas, 
encouraged by progress in the theory of rotating Kerr-Newman 
black holes \cite{burinskii00} or toroidal magnetic 
fields in helical plasma flows \cite{lortz03}. Here,
looking at the problem from a completely 
different point of view
we have unravelled a similar solenoidal structure
whose properties match surprisingly well those of the electron.
Since our model is pretty much in line with both old and 
modern ideas about the electron's structure, our identification of 
the triplet and hexaplet with, respectively, the electron and 
its neutrino is entirely natural and logical.

We have seen that a classical model with non-linear fields (\ref{eq:sfield0})
can reproduce quantum properties of the electron. 
Using the symmetry of these fields we have been able to uncover 
the structure formation and symmetry-breaking mechanisms, which are probably
responsible for the formation of the entire diversity of elementary particles. 
By deriving their quantum properties in a classical way our model supports 
the assertion that quantum mechanics could, indeed, arise from classical 
(albeit non-deterministic) processes. The impossibility within our framework
to deal with isolated systems (Sect.\,3) coheres with the ideas of A.\,Land\'e who
maintained  \cite{lande65} that uncertainty is a physical principle equally important 
for both classical and quantum physics. 
The symmetry of time-reversibility is also 
broken in our model because, as is known \cite{prigogine79}, even a negligibly small 
noise in a deterministic N-body system can cause the loss of reversibility.

The framework outlined here opens a few promising directions for further research. 
First, it is of importance to specify the parameters of interaction between the colour
particles under discussion. The basic field does not necessarily have to be of the 
proposed simplest form (\ref{eq:sfield0}). Its main feature (the capability 
of generating equilibrium particle configurations) could be derived 
from various physical considerations, one of which could be a detailed
analysis of the collective behaviour of autosolitons in appropriate media,
likely leading to the desirable shape of the potential.

The next step would be the study of the behaviour of the nine-body
system -- triplet identified here with the electron -- under different background
potentials and initial energy conditions in order to derive the exact
value of the gyromagnetic ratio for this system. 
Then it would be logical to calculate the interaction potential
between the triplet and hexaplet systems, the bound state of which 
is expected to be a highly nonlinear oscillator with one of its semiamplitudes
growing at the expense of the other. Calculating the average disruption
time and some other parameters of this oscillator is a challenging task, likely
leading to unexpected ramifications and new results.



\begin{thebibliography}{999}
%

%
\bibitem{landau81}
L.\,D.\,Landau, E.\,M.\,Lifshitz, Quantum Mechanics: Non-relativistic Theory 
(introduction), 3rd ed., Butterworth-Heinemann, Oxford, England, 1981.
%
\bibitem{penrose98}
R.\,Penrose, Quantum computation, entanglement and state reduction,
Phil. Trans. R. Soc. Lond. A 356 (1998) 1927--1939.
%
\bibitem{elitzur05}
A.\,C.\,Elitzur, S.\,Dolev, N.\,Kolenda (eds.): Quo Vadis Quantum Mechanics?
Springer, Berlin-Heidelberg, 2005;\,
%
D.\,Dragoman, M.\,Dragoman, Quantum-Classical Analogies,
Springer, Berlin-Heidelberg, 2004.
%
\bibitem{prigogine62}
I.\,Prigogine, Non-equilibrium Statistical Mechanics, 
Wiley-Interscience Publ., New York, 1962.
%
\bibitem{strocchi66}
F.\,Strocchi, Complex coordinates and quantum mechanics, 
Rev. Mod. Phys. 38 (1966) 36--40;\,
%
A.\,Heslot, Quantum mechanics as a classical theory,
Phys. Rev. D 31 (1985) 1341--1348.
%
\bibitem{joos03}
E.\,Joos, H.\,D.\,Zeh, C.\,Kiefer, C.\,Giulini, K.\,Kupsch and 
I.-O.\,Stamatescu, Decoherence and the Appearance of a Classical
World in Quantum Theory, 2nd ed., Springer, Berlin-Heidelberg, 2003.
%
\bibitem{almeida05}
V.\,Almeida, D.\,Peralta-Salas and M.\,Romera, Can two chaotic systems give rise 
to order? Physica D 200 (2005) 124--132;\,
%
G.\,A.\,Gottwald, I.\,Melbourne, Testing for chaos in deterministic systems
with noise, Physica D 212 (2005) 100--110.
%
\bibitem{hooft01a}
G.\,'t~Hooft, Determinism in free bosons, 
Int. J. Theor. Phys. 42 (2003) 355--361;\,
%
G.\,'t~Hooft, Quantum Mechanics and Determinism,
e-print arxiv.org/hep-th/0105105, 2001;\,
%
G.\,'t~Hooft, How does God plays dice?
e-print arxiv.org/hep-th/0104219, 2001.
%
\bibitem{prezhdo02}
O.\,V.\,Prezhdo, Classical mapping of second-order quantized Hamiltonian
dynamics, J. Chem. Phys. 117 (2002) 2995--3002.
%
\bibitem{keshavamurthy97}
S.\,Keshavamurthy, G.\,S.\,Ezra, 
Eigenstate assignments and the quantum-classical correspondence 
for highly-excited vibrational states of the Baggot H$_2$O 
Hamiltonian, J. Chem. Phys. 107 (1997) 156--179;\,
%
F.\,M.\,Izrailev, Quantum-classical correspondence for isolated 
systems of interacting particles: Localization and ergodicity 
in energy space, Physica Scripta T 90 (2001) 95--104;\,
%
C.\,H.\,Lewenkopf, R.\,O.\,Vallejos, Classical-quantum correspondence 
for the scattering dwell time, Phys. Rev. E 70 (2004) 036214;\,
%
G.\,Reinisch, Classical Hamiltonian description of a two-level 
quantum system: The quantum Zeno effect, Physica D
119 (1999) 239--249;\, 
%
A.\,I.\,Pesci, R.\,E.\,Goldstein and H.\,Uys, Mapping of the classical kinetic 
balance equations onto the Schr\"odinger equation, Nonlinearity 18 (2005) 211--226.
%
\bibitem{hooft97}
G.\,'t~Hooft, Quantummechanical behaviour of a deterministic model,
Found. Phys. Lett. 10 (1997) 105--112;\,
%
G.\,'t~Hooft, Quantum gravity as a dissipative deterministic system,
 Class. Quantum Grav. 16 (1999) 3263--3279.
%
\bibitem{elze05}
H.-T.\,Elze, Determinism and a supersymmetric classical model of
quantum fields, Brasilian J. Phys. 35 (2005) 343--350;\,
%
E.\,Gozzi and M.\,Reuter, Hidden BRS invariance in classical mechanics II,
Phys. Rev. D 40 (1989) 3363--3377;\,
%
L.\,E.,Reichl, The Transition to Chaos in Conservative Classical Systems:
Quantum Manifestation, Springer, Berlin, 1992;\,
%
E.\,Gozzi, M.\,Reuter and W.\,D.\,Thacker, Symmetries of the classical path 
integral on a generalized phase-space manifold, 
Phys. Rev. D 46 (1992) 757--765.
%
\bibitem{durr92}
D.\,D\"urr, S.\,Goldstein and N.\,Zangh\`i, Quantum equilibrium 
and the origin of absolute uncertainty, J. Stat. Phys. 67 (1992) 843--907;\,
%
P.\,S.\,Wesson, Space-time uncertainty from higher-dimensional determinism,
Gen. Rel. Grav. 36 (2004) 451--457;\,
%
E.\,Fredkin, An information process based on reversible cellular automata,
Physica D 45 (1990) 254--270.
\bibitem{prigogine01}
I.\,Prigogine, T.\,Petrovsky and G.\,Ordonez, Time symmetry breaking 
and stochasticity in Hamiltonian physics, in V.A.Petrov (ed) Proc. 24-th
IHEP Workshop, IHEP Press, Protvino, 2001, pp. 204--220;\,
%
C.\,P.\,Sun, X.\,F.\,Liu, and S.\,X.\,Yu, Quotend Construction of 't\,Hoofts
Quantum Equivalence Classes, e-print arxiv.org/hep-th/0006105, 2000.
%
\bibitem{griffiths87} 
D.\,Griffiths, Introduction to Elementary
Particles, Wiley and Sons, Chichester, 1987.
%
\bibitem{dirac39} 
P.\,A.\,M.\,Dirac, Theory of radiating electrons, 
Proc. Roy. Soc. A 167 (1938) 148--169.
%
\bibitem{bernabeu00}
J.\,Bernab\'eu , L.\,G.\,Cabral-Rosetti, J.\,Papavassiliou and J.\,Vidal,
Charge radius of the neutrino, Phys. Rev. D 62 (2000) 113012;\,
%
E.\,A.\,Kuraev, L.\,N.\,Lipatov and T.\,V.\,Shishkina, QED radiative corrections 
to impact factors, J. Exp. Theor. Phys. 92 (2001) 203--209;\,
%
L.\,L.\,Foldy, Neutron-electron interaction, Rev. Mod. Phys. 30 (1958) 471--481. 
%
\bibitem{kerner94}
B.\,S.\,Kerner, V.\,V.\,Osipov, Autosolitons: A New Approach to Problems
of Self-Organization and Turbulence, Kluwer, Dordrecht, 1994. 
%
\bibitem{wheeler62}
J.\,A.\,Wheeler, Geometrodynamics, Academic Press, New York, 1962.
%
\bibitem{bohm80}
D.\,Bohm, Wholeness and the Implicate Order, Rutledge, London, 1980;\,
%
H.\,I.\,Ringermacher, L.\,R.\,Mead, Induced matter: Curved N-manifolds
encapsulated in Riemann-flat N+1 dimensional space, J. Math. Phys. 46 
(2005) 102501;\,
%
J.\,Ambjorn, J.\,Jurkewicz, R.\,Loll, The universe from scratch,
Contemp. Phys. 47 (2006) 102--117.
%
%
\bibitem{ritis83}
R.\,de\,Ritis, M.\,Lavorgna, G.\,Platania, C.\,Stornaiolo, 
Spin fluid in Einstein-Cartan theory: A variational principle and an
extension of the velocity potential representation, Phys. Rev. 
D 28 (1983) 713--717;\,
%
C.\,Sivaram, Fundamental interactions in the early universe, 
Intern. J. Theor. Phys. 33 (1994) 2407--2413.
%
\bibitem{dzhunushaliev98}
V.\,D.\,Dzhunushaliev, Spherically-symmetric solution for torsion
and the Dirac equation in 5D spacetime, Intern. J. Mod. Phys. D 7 (1998) 
909--915;\,
%
M.\,Alimohammadi, Some correlators of SU(3)$_3$ WZW models on higher-genus
Riemann surfaces, Mod. Phys. Lett. A 9 (1994) 381--398.
%
%
\bibitem{jackiw00}
R.\,Jackiw, V.\,P.\,Nair and S.-Y.\,Pi, Chern-Simons reduction and 
non-Abelian fluid mechanics, Phys. Rev. D 62 (2000) 085018;\,
%
B.\,Bystrovic, R.\,Jackiw, H.\,Li, V.\,P.\,Nair and S.-Y.\,Pi,
  Non-Abelian fluid dynamics in Lagrangian formulation,
Phys. Rev. D 67 (2003) 025013.
%
\bibitem{dai06}
J.\,Dai and V.\,P.Nair, Color Skyrmions in the quark-gluon plasma,
Phys. Rev. D 74 (2006) 085014.
%
\bibitem{osenda02}
O.\,Osenda, P.\,Serra, F.\,A.\,Tamarit, Non-equilibrium properties 
of small Lennard-Jones clusters, Physica D 168--169 (2002) 336--340.
\bibitem{pikovsky03}
%
A.\,Pikovsky, M.\,Rosenblum, J.\,Kurths, Synchronization: A Universal
Concept in Nonlinear Sciences, Cambridge Univ. Press, Cambridge, 2003;\,
%
H.\,F.\,El-Nashar, Y.\,Zhang, H.\,A.\,Cerdeira and F.\,Ibiyinka A,
Synchronization in a chain of nearest neighbors coupled oscillators with 
fixed ends, Chaos 13 (2003) 1216--1225;\,
%
W.\,Qin, G.\,Chen, Coupling schemes for cluster synchronization in
coupled Josephson equations, Physica D 197 (2004) 375--391. 
\bibitem{yershov05} V.\,N.\,Yershov, Equilibrium configurations of tripolar 
charges, Few-Body Syst. 37 (2005) 79--106.
%
\bibitem{brizhik06}
L.\,S.\,Brizhik, A.\,A.\,Eremko, B.\,Piette, W.\,Zakrzewski, 
Electron self-trapping on a nanocircle, Physica D 218 (2006) 36--55. 
\bibitem{baroni98}
L.\,Baroni, A.\,Cuccoli, V.\,Tognetti, R.\,Vaia, Quantum effects on the localization
of a particle in a double-well potential, Physica D 117 (1998) 375--378;\,
%
A.\,Igarashi, H.\,S.\,Yamada, Quantum dynamics and delocalization in coherently
driven one-dimensional double-well system, Physica D 221 (2006) 146--156.
\bibitem{parker78} 
E.\,N.\,Parker, Cosmical Magnetic Fields, Clarendon Press, New York, 1979;\,
%
S.\,M.\,Cox, P.\,C.\,Matthews, New instabilities in two-dimensional
rotating convection and magnetoconvection, Physica D 149 (2001) 210--229. 
%
\bibitem{pipin00} 
V.\,V.\,Pipin, L.\,L.\,Kichatinov, The solar dynamo and integrated 
irradiance variations in the course of 
the 11-year cycle, Astron. Rep. 44 (2000) 771--779;\,
%
W.\,Liu, G.\,Haller, Inertial manifolds and completeness of eigenmodes
for unsteady magnetic dynamos, Physica D 194 (2004) 297--319. 
\bibitem{kadomtsev93}
B.\,B.\,Kadomtsev, in E.\,W.\,Laing (ed.) Tokamak Plasma: A Complex 
Physical System, Inst. of Phys. Publ., Bristol, Philadelphia, 1993;\,
%
J.\,A.\,Wesson, Tokamaks, 2nd ed., Oxford Univ. Press, Oxford, 1997;\, 
%
R.\,B.\,White, The Theory of Toroidally Confined Plasmas,
2nd ed., Imperial College Press, London, 2001.
%
\bibitem{kleman83} 
M.\,Kleman, Lines and Walls, Wiley and Sons, Chichester, 1983.
%
\bibitem{wigner34}
E.\,Wigner, On the interaction of electrons in metals, 
Phys. Rev. 46 (1934) 1002--1011.  
\bibitem{maksym96}
P.\,A.\,Maksym, Eckardt frame theory of interacting electrons
in quantum dots, Phys. Rev. B 53 (1996) 10871--10886;\, 
%
H.-M.\,M\"uller, S.\,E.\,Koonin, Phase transitions in quantum dots,
Phys. Rev. B 54 (1996) 14532--14539;\, 
%
R.\,Egger, W.\,H\"ausler, C.\,H.\,Mak and H.\,Grabert, Crossover from 
Fermi liquid to Wigner molecule behaviour in quantum dots,
Phys. Rev. Lett. 82 (1999) 3320--3323;\,
%
C.\,Yannouleas and U.\,Landman, Spontaneous symmetry breaking in 
single and molecular quantum dots, Phys. Rev. Lett. 82 (1999) 5325--5328. 
%
\bibitem{walker94}
B.\,Walker et al, Precision measurement of strong field double ionization
of helium, Phys. Rev. Lett. 73 (1994) 1227--1230.
%
\bibitem{schlagheck99}
P.\,Schlagheck, A.\,Buchleitner, Stable classical configurations in 
strongly driven helium, Physica D 131 (1999) 110--124. 
\bibitem{dirac28a}
P.\,A.\,M.\,Dirac, The quantum theory of the electron,
Proc. Roy. Soc. A 117 (1928) 610--624;\,
%
Proc. Roy. Soc. A 118 (1928) 351--361.
%
\bibitem{kovachev04}
L.\,M.\,Kovachev, Vortex solutions of the nonlinear optical
Maxwell-Dirac equations, Physica D 190 (2004) 78--92.
%
\bibitem{parson15} 
A.\,L.\,Parson, A magneton theory of the structure of the atom,
Smithsonian Miscellaneous Collection {\bf 65} No.11, 
Publ. No. 2371 (1915) 1--80;\,
%
A.\,Land\'e, The coupling of electron rings and the optic 
rotating capacity of asymmetrical molecules, Phys. Zeitschrift
19 (1918) 500--505;\,
%
H.\,S.\,Allen, The case for a ring-electron,
Proc. Phys. Soc. London 31 (1919) 49--68.
%
\bibitem{compton19} 
A.\,Compton, The size and shape of the electron,
Phys. Rev. 14 (1919) 247--259.
%
\bibitem{burinskii00}
A.\,Burinskii, Structure of spinning particle suggested by 
gravity, supergravity and low energy string theory, 
Czech. J. Phys. 50 Suppl. S1 (2000) 201--206.
%
\bibitem{lortz03} 
D.\,Lortz, On the structure of the electron,
Z. Naturforsch. 58 A (2003) 491--493.
%
\bibitem{lande65}
A.\,Land\'e, New Foundations of Quantum Physics, 
Cambridge Univ. Press, Cambridge, 1965.
%
\bibitem{prigogine79}
I.\,Prigogine, Irreversibility and randomness, Astroph. and Space Science
65 (1979) 371--381;\,
%
N.\,Komatsu, T.\,Abe, Numerical irreversibility in time-reversible
molecular dynamics simulation, Physica D 195 (2004) 391--397.

%
\end{thebibliography}
\end{document}